\documentclass[fleqn,usenatbib]{mnras}

\usepackage{newtxtext,newtxmath}

\usepackage[T1]{fontenc}

\DeclareRobustCommand{\VAN}[3]{#2}
\let\VANthebibliography\thebibliography
\def\thebibliography{\DeclareRobustCommand{\VAN}[3]{##3}\VANthebibliography}

\usepackage{graphicx}	
\usepackage{amsmath}	

\title[Magnetic fields in cosmic web filaments] {Magnetic field strength in cosmic web filaments}

\author[E.~Carretti et al.]{
Ettore~Carretti,$^{1}$\thanks{E-mail: carretti@ira.inaf.it (EC)}
V.~Vacca,$^{2}$
S.~P.~O'Sullivan,$^{3}$
G.~H.~Heald,$^{4}$
C. Horellou,$^{5}$ 
H.~J.~A.~R\"ottgering,$^{6}$ \and
A.~M.~M.~Scaife,$^{7,8}$
T.~W.~Shimwell,$^{9,6}$
A.~Shulevski,$^{6}$ 
C.~Stuardi,$^{10,1}$
T.~Vernstrom$^{4}$ \\
\\
$^{1}$INAF Istituto di Radioastronomia, Via Gobetti 101, 40129 Bologna, Italy\\
$^{2}$INAF Osservatorio Astronomico di Cagliari, Via della Scienza 5, 09047 Selargius (CA), Italy\\
$^{3}$School of Physical Sciences and Centre for Astrophysics \& Relativity, Dublin City University, Glasnevin D09 W6Y4, Ireland\\
$^{4}$CSIRO, Space and Astronomy, PO Box 1130, Bentley WA 6102, Australia\\
$^{5}$Chalmers University of Technology, Dept  of Space, Earth and Environment, Onsala Space Observatory, SE-43992 Onsala, Sweden\\
$^{6}$ Leiden Observatory, Leiden University, PO Box 9513, NL-2300 RA Leiden, The Netherlands\\
$^{7}$Jodrell Bank Centre for Astrophysics, Department of Physics \& Astronomy, University of Manchester, M13 9PL, UK\\
$^{8}$The Alan Turing Institute, 96 Euston Road, London NW1 2DB, UK\\
$^{9}$ASTRON, Netherlands Institute for Radio Astronomy, Oude Hoogeveensedijk 4, 7991 PD, Dwingeloo, The Netherlands\\
$^{10}$Dipartimento di Fisica e Astronomia, Universit\'a di Bologna, via Gobetti 93/2, 40122 Bologna, Italy
}

\date{Accepted XXX. Received YYY; in original form ZZZ}

\pubyear{2021}

\begin{document}
\label{firstpage}
\pagerange{\pageref{firstpage}--\pageref{lastpage}}
\maketitle

\begin{abstract}
We used the Rotation Measure (RM) catalogue derived  from the LOFAR Two-metre Sky Survey Data Release 2 (LoTSS DR2) at 144-MHz to measure the evolution with redshift  of the extragalactic RM (RRM: Residual RM) and the polarization fraction ($p$) of sources in low density environments. We also measured  the same at 1.4-GHz by cross-matching with the NRAO VLA Sky Survey RM catalogue. We find that RRM versus redshift is flat  at 144-MHz, but, once redshift-corrected, it shows evolution at high significance. Also $p$ evolves with redshift with a decrement by a factor of $\sim$8 at $z\sim2$. Comparing the 144-MHz and 1.4-GHz data, we find that the observed RRM and $p$  are most likely to have an origin local to the source at 1.4-GHz, while a  cosmic web filament origin is favoured at 144-MHz. If we attribute the entire signal to   filaments, we infer a mean rest frame  RRM  per filament of RRM$_{0,f} = 0.71 \pm 0.07  \,\, \rm rad\,m^{-2}$ and a magnetic field per filament of $B_f = 32 \pm 3$ nG. This is in agreement with estimates obtained with a complementary method based on synchrotron emission stacking, and with cosmological simulations if primordial magnetic fields are  amplified by  astrophysical source field seeding. The measurement  of an RRM$_{0,f}$  supports the presence of diffuse baryonic gas in filaments. We also  estimated a  conservative upper limit of the filament magnetic turbulence of   $\sigma_{\rm RRM_{0,f}} =0.039 \pm 0.001 \,\, \rm rad\,m^{-2}$, concluding  that the ordered magnetic field component dominates in filaments. 
\end{abstract}




\section{Introduction}
Measuring the evolution of the cosmic magnetic field with cosmic time  helps to  understand its genesis from primordial fields and whether field seeding and amplification  by astrophysical sources has played a role \citep[e.g.,][]{2011ApJ...738..134A, 2015A&A...580A.119V, 2016RPPh...79g6901S, 2017CQGra..34w4001V, 2020MNRAS.495.2607O, 2021MNRAS.505.5038A}. Cosmic web filaments are ideal for this purpose, since  matter and fields are less processed and closer to the initial conditions. Furthermore, simulations predict that the intensity of filament magnetic fields  can help discriminate  between the possible scenarios that have magnetised these cosmic structures \citep[e.g.,][]{2015A&A...580A.119V}.
The time evolution can also inform us about the evolution of extragalactic sources themselves, such as a change in the physical conditions of the source and its environment \citep[e.g.,][]{2008ApJ...676...70K, 2021A&A...653A.155B}. 

 Effective ways to study the magnetic field evolution with time are  the behaviour with redshift $z$ of the Rotation Measure (RM) and the polarization fraction of extragalactic sources. The polarization angle $\phi$ of linearly polarized radiation travelling through a magnetised, ionised gas is rotated by
\begin{equation}
    \Delta \phi = {\rm RM} \, \lambda^2 
\end{equation}
at wavelength $\lambda$. RM is: 
\begin{equation}
    {\rm RM} = 0.812\int_z^0 \frac{n_e(z')B_\parallel (z')}{(1+z')^2} \,\frac{dl}{dz'} \, dz'
    \label{eq:rm}
\end{equation}
where $z$ is the source redshift, the integration is performed from the source to the observer along the path length $l$ (pc), $n_e$ is the electron density (cm$^{-3}$), and $B_\parallel$ is the magnetic field along the line of sight ($\mu$G). Hence, RM bears information on the magnetised medium the radiation travels through and can be used to trace the evolution of the magnetic field  \citep[e.g.,][]{2008ApJ...676...70K}. 
The polarization fraction evolution can be related to a change of depolarization \citep{1998MNRAS.299..189S} and in turn a change of magneto-ionic physical conditions at the source \citep[e.g.,][]{2021A&A...653A.155B} or in the  intergalactic medium (IGM).

The behaviour of the extragalactic source RM with redshift was investigated in the past decades, finding no evolution \citep{1971PThPS..49..181F,1972A&A....19..104R,1977A&A....61..771K,1979PASJ...31..125S,1982ApJ...263..518K,1982MNRAS.201..365T,1984ApJ...279...19W, 1995ApJ...445..624O, 2003AcASn..44S.155Y,2008Natur.454..302B, 2018MNRAS.475.1736V,  2020PASA...37...29R}. However, using samples of a few hundred sources,  \cite{2008ApJ...676...70K} and \citet{2016ApJ...829....5L} found  a hint of evolution at low significance, at GHz frequencies, which was attributed to RM originating local to the source \citep{2008ApJ...676...70K}. This evolution was not confirmed using a much larger sample of 4003 sources \citep{2012arXiv1209.1438H}, making the results inconclusive so far. The evolution of RM with redshift has also been used to investigate the magnetic field of the intracluster medium in galaxy clusters using the differential RM of physical source pairs \citep{2021arXiv211201763X}.

No clear evolution of fractional polarization was found by \citet{2012arXiv1209.1438H} and \citet{2016ApJ...829....5L}, while \citet{2021A&A...653A.155B} recently found an anti-correlation  in a low-brightness 
sample of 56 sources at 1.4-GHz that was attributed to evolution of the environment local to the source. 

The detection of magnetic fields in filaments has been the subject of intense research in recent years \citep{2020Galax...8...53H}. Several upper limits were  set employing a number of different approaches \citep{2017MNRAS.467.4914V, 2017MNRAS.468.4246B, 2018MNRAS.479..776V, 2019A&A...622A..16O,2019ApJ...878...92V, 2020MNRAS.495.2607O, 2021A&A...652A..80L} 
and a first detection was obtained stacking the synchrotron emission from bridges connecting galaxy clusters \citep{2021MNRAS.505.4178V}.
A direct detection of a filament between a close pair of interacting  galaxy clusters  was also obtained establishing the presence of magnetic fields in the IGM \citep{2019Sci...364..981G}.
Cosmological simulations have been run to study the conditions required to generate magnetic fields in filaments that range from a few nG to a few tens of nG depending on whether only primordial fields  or additional  astrophysical source seeding are involved \citep{2010ApJ...723..476A,2011ApJ...738..134A,2015MNRAS.453.1164G,  2015A&A...580A.119V, 2017CQGra..34w4001V, 2020MNRAS.495.2607O, 2021MNRAS.505.5038A}. 

In this work, conducted within the LOFAR Magnetism Key Science Project\footnote{https://lofar-mksp.org/} (MKSP), we compute and analyse the behaviour with redshift of the extragalactic source RM and fractional polarization of a low frequency (144-MHz) RM catalogue (O'Sullivan et al., in prep.).  This catalogue was derived from Stokes $Q$ and $U$ data cubes of the LOFAR Two-metre Sky Survey Data Release 2 \citep[(LoTSS DR2, Shimwell et al., in prep, ][]{2017A&A...598A.104S,2019A&A...622A...1S} in a collaborative effort between the LOFAR Surveys Key Science Project\footnote{https://lofar-surveys.org/} and the MKSP.  At this low frequency the polarized radiation can survive depolarization only if it is emitted and propagates through low density environments \citep{2019A&A...622A..16O,2020A&A...638A..48S,2020MNRAS.495.2607O} and our analysis allows us  to investigate the evolution of magnetism in such environments. We find that   an origin of RRM and $p$ in cosmic web filaments is favoured and that enables us to derive  properties of the magnetic field  in cosmic filaments such as intensity and turbulence.  

The paper is organised as follows. Section \ref{sec:cat} describes the RM catalogues used in the analysis.  Section \ref{sec:evo} computes the behaviours with redshift and other related quantities at 144-MHz and 1.4-GHz. Section \ref{sec:discussion} discusses the results and  possible scenarios for the origin of the RM and fractional polarization, and in Section \ref{sec:conc} we draw our conclusions. 

Throughout the paper, we assume a $\Lambda$CDM cosmological model with $H_0 = 67.4$-km~s$^{-1}$~Mpc$^{-1}$, $\Omega_M = 0.315$, and $\Omega_\Lambda =0.685$ \citep{2020A&A...641A...6P}. We also use the term $h=H_0 /100$-km~s$^{-1}$~Mpc$^{-1}$.  

\section{RM Catalogues}
\label{sec:cat}

\subsection{LoTSS DR2}
Our analysis is based on the RM catalogue derived from Stokes $Q$ and $U$ data cubes of the LoTSS DR2 survey  (O'Sullivan et al. in prep.). Here we report the main catalogue features relevant to this work.   It consists of 2461 RMs detected at a central frequency of 144-MHz,  bandwidth of 48-MHz, channel bandwidth of 97.6-kHz, angular resolution of 20-arcsec, over 5720-deg$^2$, obtained using the method of RM-synthesis \citep{2005A&A...441.1217B}. The RM error budget  is dominated by random ionospheric RM correction residuals, which  are $\sim 0.05$-rad~m$^{-2}$ (O'Sullivan et al. in prep.). A systematic term as large as 0.1--0.3-rad m$^{-2}$ is also present, again related to ionospheric RM correction residuals \citep{2013A&A...552A..58S,2019MNRAS.483.4100P}. A total number of 1949 sources had a positive cross-match with redshift catalogues, 1046 of which are spectroscopic redshifts. 

Photometric redshifts of the identified sources have a median error of $\sigma_{z,\rm  phot} \sim 0.1$, comparable to the redshift bin width used here, and we excluded them keeping sources with spectroscopic redshift only.   A Galactic cut excluding sources at $|b|  <  25^\circ$ was  applied to exclude the region with highest Galactic RM values, giving  1016 sources. 
   \begin{figure}
   \centering
    \includegraphics[width=\columnwidth]{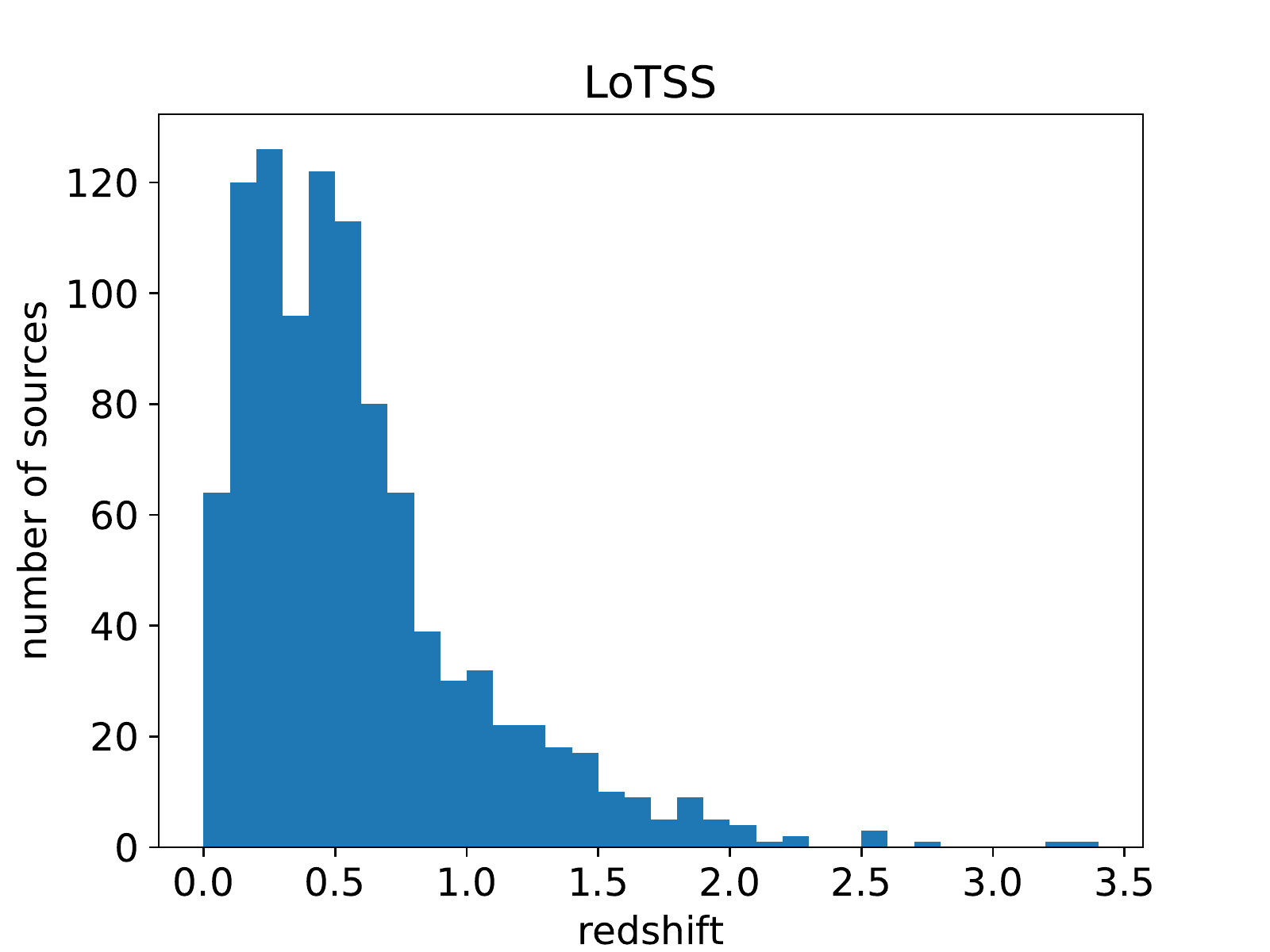}
   \caption{Distribution with redshift of the sources used from the LOFAR LoTSS DR2 RM catalogue.}
              \label{Fig:zdist}%
    \end{figure}

The source distribution  with redshift is shown in Figure~\ref{Fig:zdist}. The median redshift is $\sim 0.5$.  Only a handful of sources have redshift $z > 2$ and we limited our  analysis to $z < 2$, for   our final sample of 1003 objects.

\subsection{NVSS}\label{sec:nvss}
We also compared the results from the LoTSS RM catalogue with those at higher frequency (1.4-GHz) from the NRAO VLA Sky Survey (NVSS) RM catalogue.  The NVSS RM catalogue \citep{2009ApJ...702.1230T} measured the RM of 37543 sources with two  narrow bands, 42-MHz wide each, centred at 1365-MHz and 1435-MHz, and  at an angular resolution of 45-arcsec. It covers all Declinations $\geq -40^\circ$.   \citet{2012arXiv1209.1438H} cross-matched it with a number of redshift catalogues obtaining  4003 matches. All redshifts are spectroscopic.  To match the selection criteria applied to the LoTSS RM catalogue, we also restricted our NVSS RM sample to sources at Galactic latitudes $|b| > 25^\circ$, resulting in a sample of 3406 sources, that  reduce  to 3055  at the redshift limit of $z < 2$. 

\citet{2019ApJ...878...92V} found that RM errors of \citet{2009ApJ...702.1230T} are  overestimated for at least part of the sample and  we recompute them following their  Equation~(19): 
\begin{equation}
    \sigma_{\rm RM_{NVSS}} = 150\,\frac{\sqrt{2}\,\, \sigma_P}{P} \,\,\rm rad/m^2
\end{equation}
with $P$ the polarized intensity of the source,  $\sigma_P$ its error, and 150 a coefficient related to the wavelengths of the two NVSS frequency bands.  

\section{Redshift evolution analysis}\label{sec:evo}

\subsection{LoTSS}\label{sec:rmvz_LoTSS}
The measured RM  is the combination of Galactic (GRM), extragalactic, and noise  components, where the extragalactic term is    either local to the source or from the foreground IGM  between the source and the Milky Way, including filaments of the cosmic web: 
\begin{equation}
    {\rm RM = GRM + RM_{local} + RM_{IGM} + RM_{noise}}
\end{equation}
A key point is that the first term has to be subtracted off to be left with the extragalactic component only (and noise), which we call the Residual RM (RRM)
\begin{equation}
    {\rm RRM = RM - GRM}. 
\end{equation} 
We estimate the GRM from~\citet{2022A&A...657A..43H}, who inferred a map of the Galactic RM from a suite of extragalactic source RM catalogues, including those from LoTSS and NVSS. This is a sophisticated evolution of former estimates of GRM maps (e.g., \citealt{2015A&A...575A.118O}) with improved errors, resolution, and sampling of the parent catalogue collection of extragalactic source RMs.

Since the extragalactic RRMs are  generally distributed around zero, an estimate of the typical RRM intensity of a sample is the rms deviation $\left< {\rm RRM}^2\right>^{1/2}$. 
For the LoTSS sample, 
if we subtract the GRM contribution as estimated straight from the \citet{2022A&A...657A..43H}  map at the exact source position, we get an rms deviation of  RRM of $\left< {\rm RRM}^2\right>^{1/2} = 0.52$-rad~m$^{-2}$ (excluding 2-sigma outliers), that further drops to  0.15-rad~m$^{-2}$ if  estimated  with  the median absolute deviation (MAD) statistics that is more robust against outliers. This is in contrast with the mean GRM error over all sources of $0.79$-rad~m$^{-2}$. The measured RRM rms is $\left< {\rm RRM}^2\right>^{1/2}= \sqrt{ \left< {\rm  RRM^2_{source}}\right> + \left< {\rm RRM^2_{noise}} \right>}$.  The noise term has to be quadratically subtracted off to be left with the source term and a noise larger than the measured term is unexpected. This is possibly because our sample is part of the catalogue suite  used to infer the GRM map and the  GRM at the exact source position might be slightly biased towards the source RM, which gives an oversubtraction \citep[the possible presence of extragalactic residuals in the GRM map is  also mentioned by ][]{2022A&A...657A..43H}. 
Actually, an inspection of the GRM map shows that it can have  a slight bump at our source positions compared to the immediate surrounding fields.  

\begin{figure}
   \centering
    \includegraphics[width=\columnwidth]{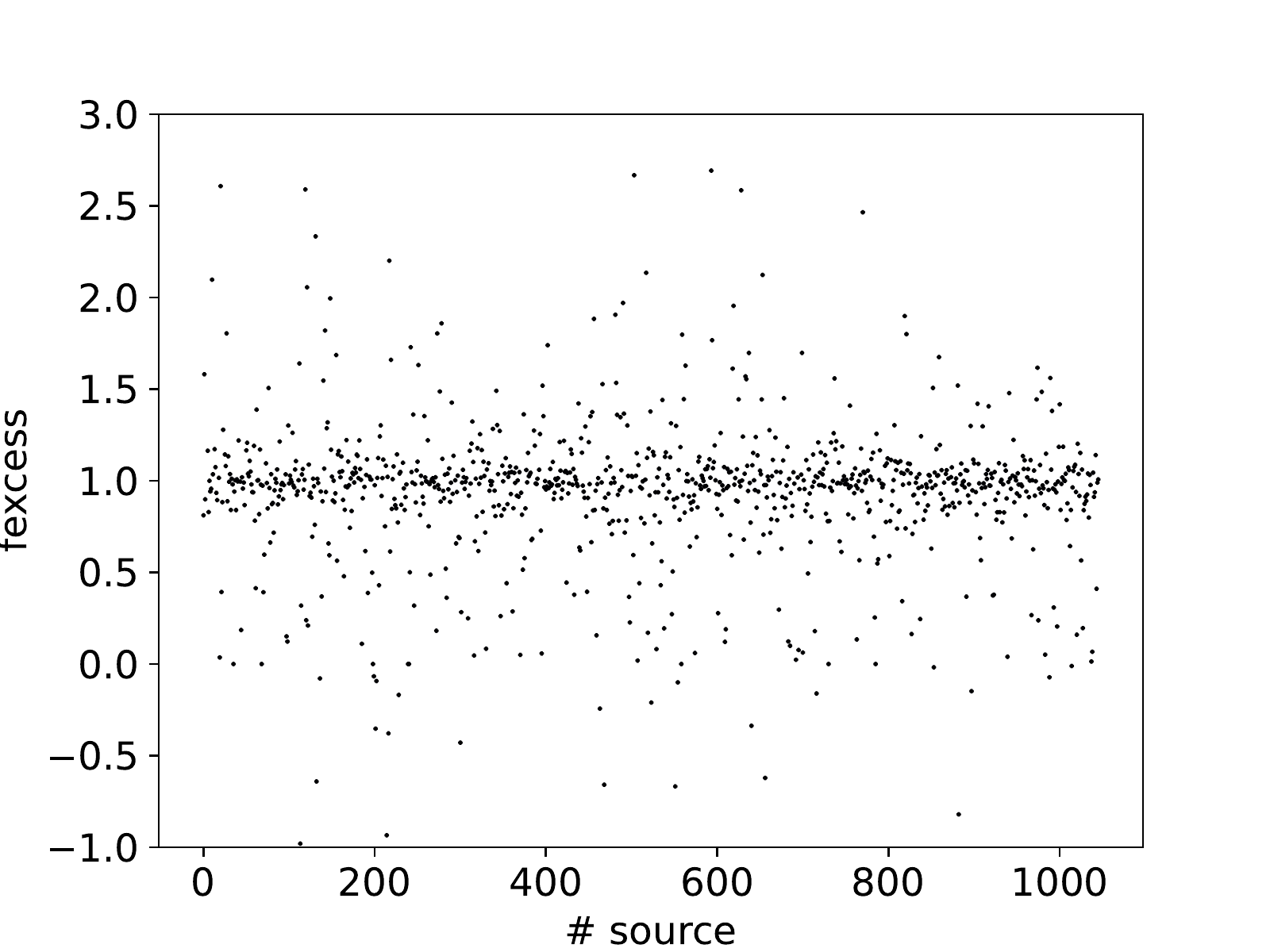}
   \caption{Fractional excess $f_{\rm excess}$ (Eqn.~\ref{eqn:fexcess}) of each individual source of the selected LoTSS DR2 RM catalogue sample.}
              \label{Fig:fex}%
    \end{figure}

To test this further, for each source we computed the difference of RM  and GRM with a reference term. As a reference we used GRM$_{1}$, the median of the GRM map over a 1-deg diameter disc centred at the source position that is approximately the average spacing between sources in the  catalogue suite used by \citet{2022A&A...657A..43H}.  The differences are: 
\begin{eqnarray}
   {\rm RM_{excess}}  & = &\rm RM - \rm GRM_1 \\
   {\rm GRM_{excess}} & = &\rm GRM - \rm GRM_1 . 
\end{eqnarray}
Their  ratio, the fractional excess 
\begin{equation}\label{eqn:fexcess}
    f_{\rm excess} = \frac{\rm RM_{excess}}{\rm GRM_{excess}}\, ,
\end{equation}
is shown in Figure~\ref{Fig:fex}. Its median is $ \bar{f}_{\rm excess} = 0.986 \pm 0.005$ and  has a narrow deviation of  $ \sigma_{f_{\rm excess}} = 0.15$
(here we  used the median and its deviation as estimated with the MAD statistics  because they are  more robust against the obvious outliers). This  means it tends to be $\rm RM =\rm GRM$ for each individual source and  suggests that GRM actually is slightly biased towards the source RM at its exact position.

   \begin{figure}
   \centering
    \includegraphics[width=\columnwidth]{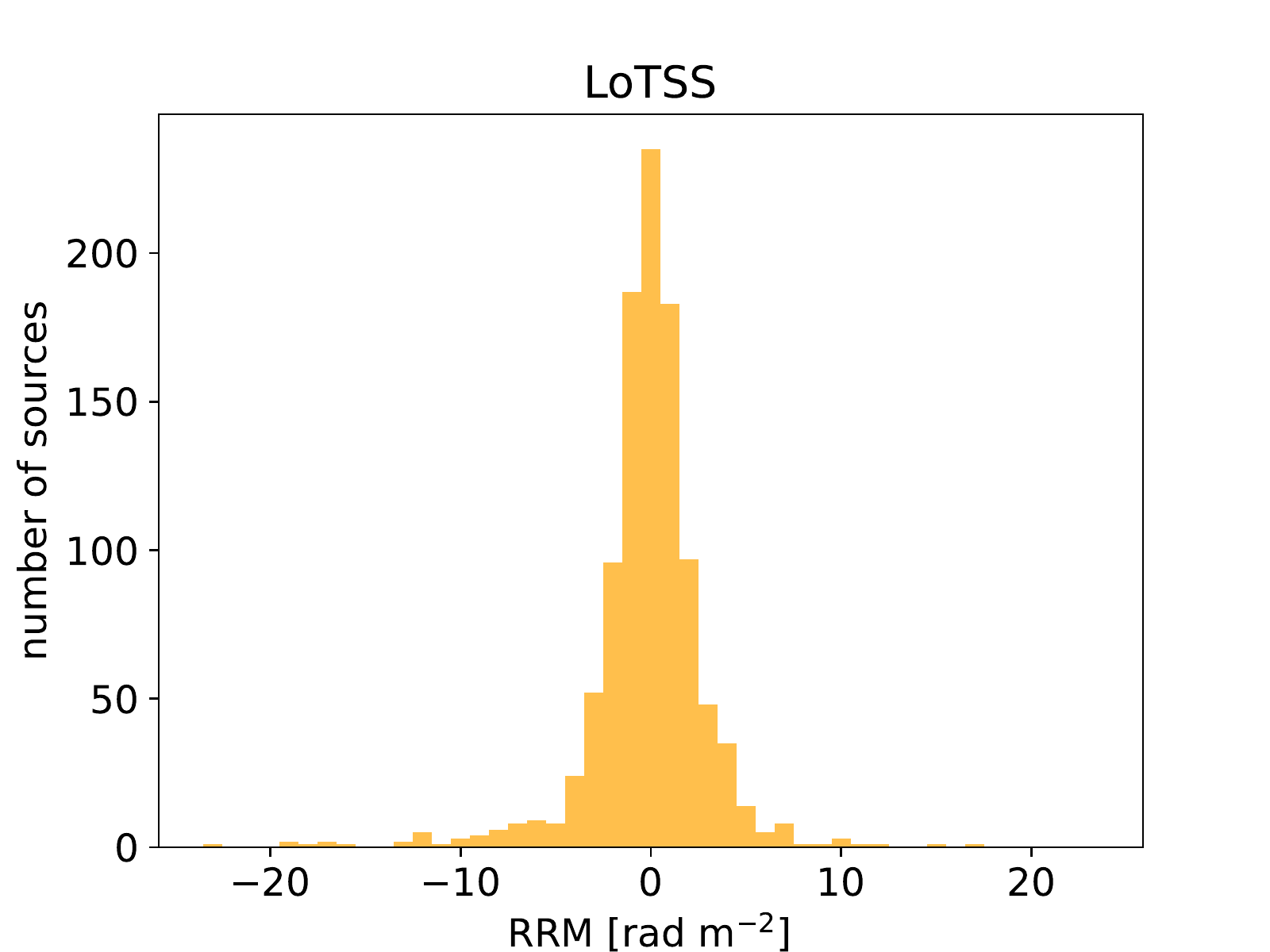}
   \caption{Distribution of RRMs of the sample from the LoTSS RM catalogue}
              \label{Fig:rrmdist}%
    \end{figure}
Instead, to estimate the GRM we have then taken $\rm GRM_1$ as defined above, because the median over a region around the source tends to mitigate the contribution of the bump at the source position  (applying a smoothing  is  also recommended by \citealt{2022A&A...657A..43H}). The distribution of the RRMs obtained with the GRMs so estimated  is shown in Figure \ref{Fig:rrmdist}. 
The RRM rms corrected for the noise is 
\begin{equation}
      \left< {\rm RRM_{LoTSS}}^2 \right>^{1/2} = 1.90\pm0.05\,\, \rm rad\, m^{-2},
      \label{eq:rrm_lotss}
\end{equation}
 where the noise terms are the RM measurement noise and the GRM$_1$ error quadratically subtracted off. 
 Because of the presence of outliers, here and throughout the paper, RRMs that are off by more than 2-sigmas were excluded. 
This result is broadly consistent with \citet{2020MNRAS.495.2607O} who estimated the  differential RM rms of random pairs at the same frequency, that is $\sqrt{2}$ times the RRM rms of a single source. Dividing their result by $\sqrt{2}$ (we used their result on   42 random pairs within 10-arcmin separation) we estimate an RRM rms of single sources of 
\begin{equation}
    \left< {\rm RRM_{rp}^2}\right>^{1/2} = 1.3\pm 0.2\,\,\rm rad\,m^{-2} 
\end{equation}
which supports that the procedure applied here is correct. 

\begin{table*}
	\centering
	\caption{RRM mean and rms of the LoTSS sample in the equal-width redshift bins $z$ (bin centre). The rms of the rest frame ${\rm RRM}_0 = C_x\,{\rm RRM}$  for the three redshift correction models $C_x$ discussed in the text are also reported.}
	\label{tab:zdisp}
	\begin{tabular}{cccccc} 
       \hline 
        z & $\rm \left< RRM \right>$ &  $\rm \left< RRM^2 \right>^{1/2}$ &  $\rm \left< (C_1\,\,RRM)^2 \right>^{1/2}$ &  $\rm \left< (C_2\,\,RRM)^2 \right>^{1/2}$  &  $\rm \left< (C_3\,\,RRM)^2 \right>^{1/2}$ \\ 
           & [rad m$^{-2}$]  & [rad m$^{-2}$]  & [rad m$^{-2}$]  & [rad m$^{-2}$]  & [rad m$^{-2}$]  \\ 
       \hline 
         0.143  & $    0.06 \pm    0.10 $   & $    1.71 \pm    0.09 $    & $    2.34 \pm    0.12 $    & $    2.01 \pm    0.10 $    & $    1.99 \pm    0.10 $ \\ 
         0.429  & $    0.07 \pm    0.11 $   & $    1.82 \pm    0.09 $    & $    3.82 \pm    0.21 $    & $    2.75 \pm    0.14 $    & $    2.63 \pm    0.13 $ \\ 
         0.715  & $    0.04 \pm    0.17 $   & $    2.25 \pm    0.14 $    & $    6.35 \pm    0.37 $    & $    4.12 \pm    0.23 $    & $    3.77 \pm    0.22 $ \\ 
         1.001  & $    0.13 \pm    0.21 $   & $    1.89 \pm    0.15 $    & $    7.52 \pm    0.62 $    & $    4.39 \pm    0.37 $    & $    3.77 \pm    0.31 $ \\ 
         1.287  & $   -0.40 \pm    0.25 $   & $    1.85 \pm    0.16 $    & $    9.64 \pm    0.90 $    & $    5.23 \pm    0.46 $    & $    4.21 \pm    0.39 $ \\ 
         1.573  & $    0.53 \pm    0.32 $   & $    1.67 \pm    0.34 $    & $   10.95 \pm    2.15 $    & $    5.63 \pm    1.09 $    & $    4.28 \pm    0.86 $ \\ 
         1.859  & $    0.09 \pm    0.39 $   & $    1.53 \pm    0.28 $    & $   12.09 \pm    2.17 $    & $    5.97 \pm    1.06 $    & $    4.30 \pm    0.76 $ \\ 
       \hline 
	\end{tabular}
\end{table*}

\begin{figure}
   \centering
    \includegraphics[width=\columnwidth]{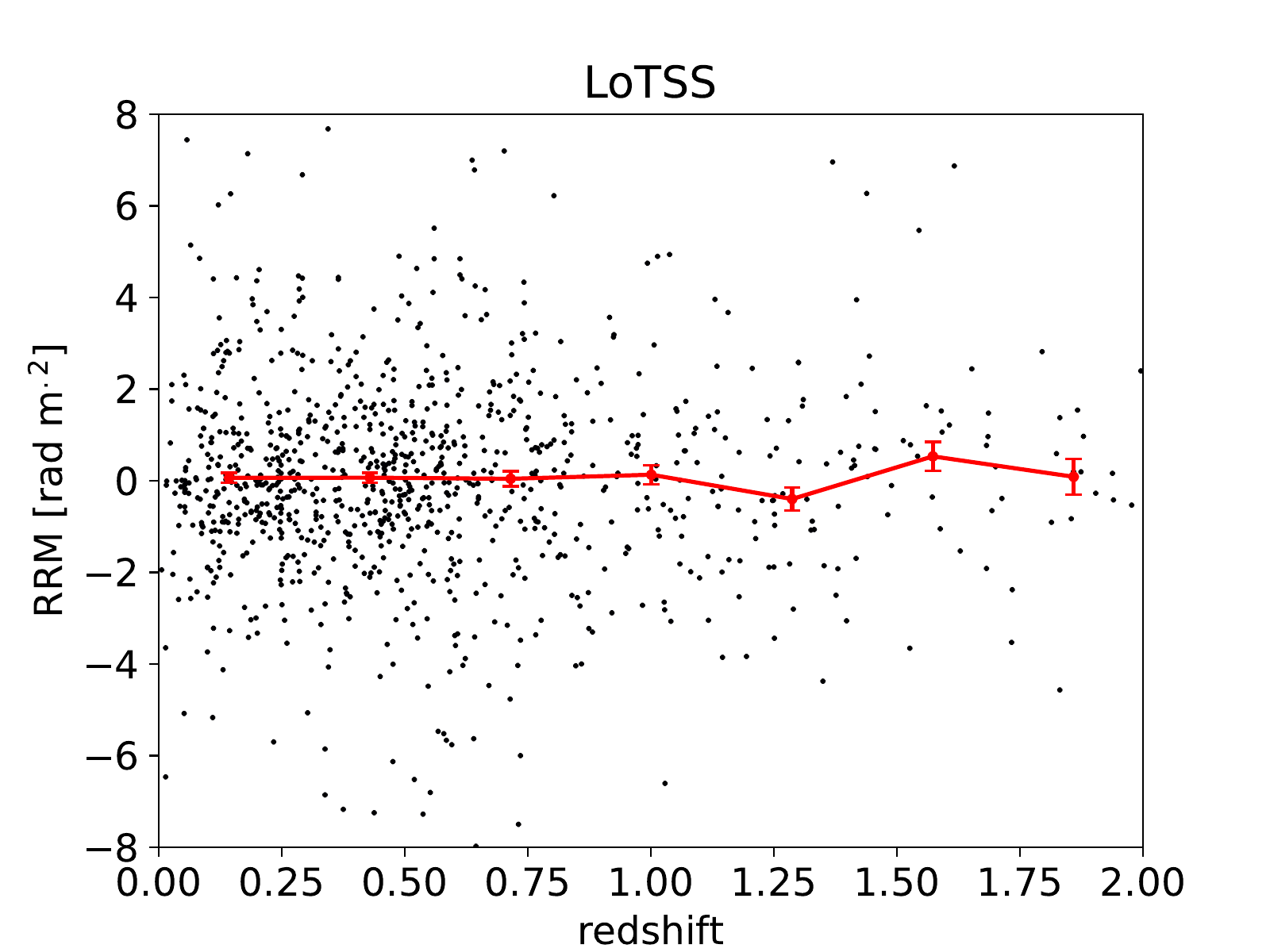}
    \includegraphics[width=\columnwidth]{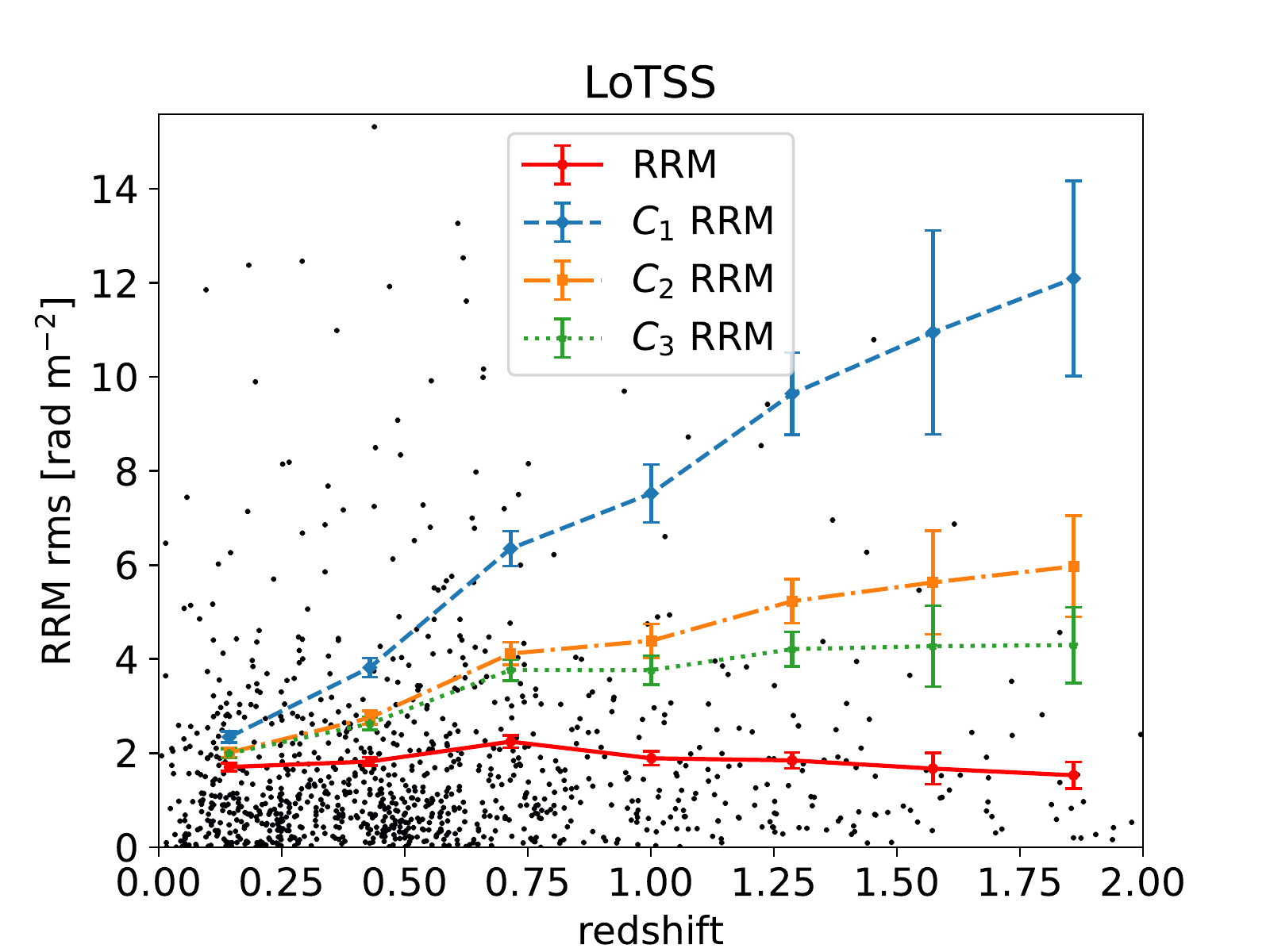}
   \caption{{\em Top}: Mean of the RRM sample selected from the LoTSS RM catalogue in equal-width redshift bins (red solid line). Individual RRMs are also shown (dots). {\em Bottom}: As for top panel but for the  RRM rms deviation  (red solid line).   The rest frame RRM$_0$ corrected for the redshift effects $C_1$ (dashed line), $C_2$ (dot-dashed line), and $C_3$ (dotted line) are also plotted. Individual |RRM|s  are  shown as dots }
              \label{Fig:zmean20}%
              \label{Fig:zdisp20}%
    \end{figure}

Figure \ref{Fig:zmean20}, top panel, and Table \ref{tab:zdisp} report the RRM mean of the LoTSS sample in equal-width redshift bins. It is zero within 2-$\sigma$ in all bins.  

Figure \ref{Fig:zdisp20}, bottom panel,  and Table \ref{tab:zdisp}  report  the RRM rms deviation   $\rm\left< RRM^2\right>^{1/2}$ in the same redshift bins (red solid line). It is flat with redshift, its linear fit gives a slope of $\beta =  -0.15 \pm 0.15$, consistent with no evolution with redshift out to $z=2$.  Here and throughout the paper, the error terms (i.e. the RM measurement noise and the GRM$_1$ error) are quadratically subtracted off from the RRM rms estimates.

   \begin{figure}
   \centering
    \includegraphics[width=\columnwidth]{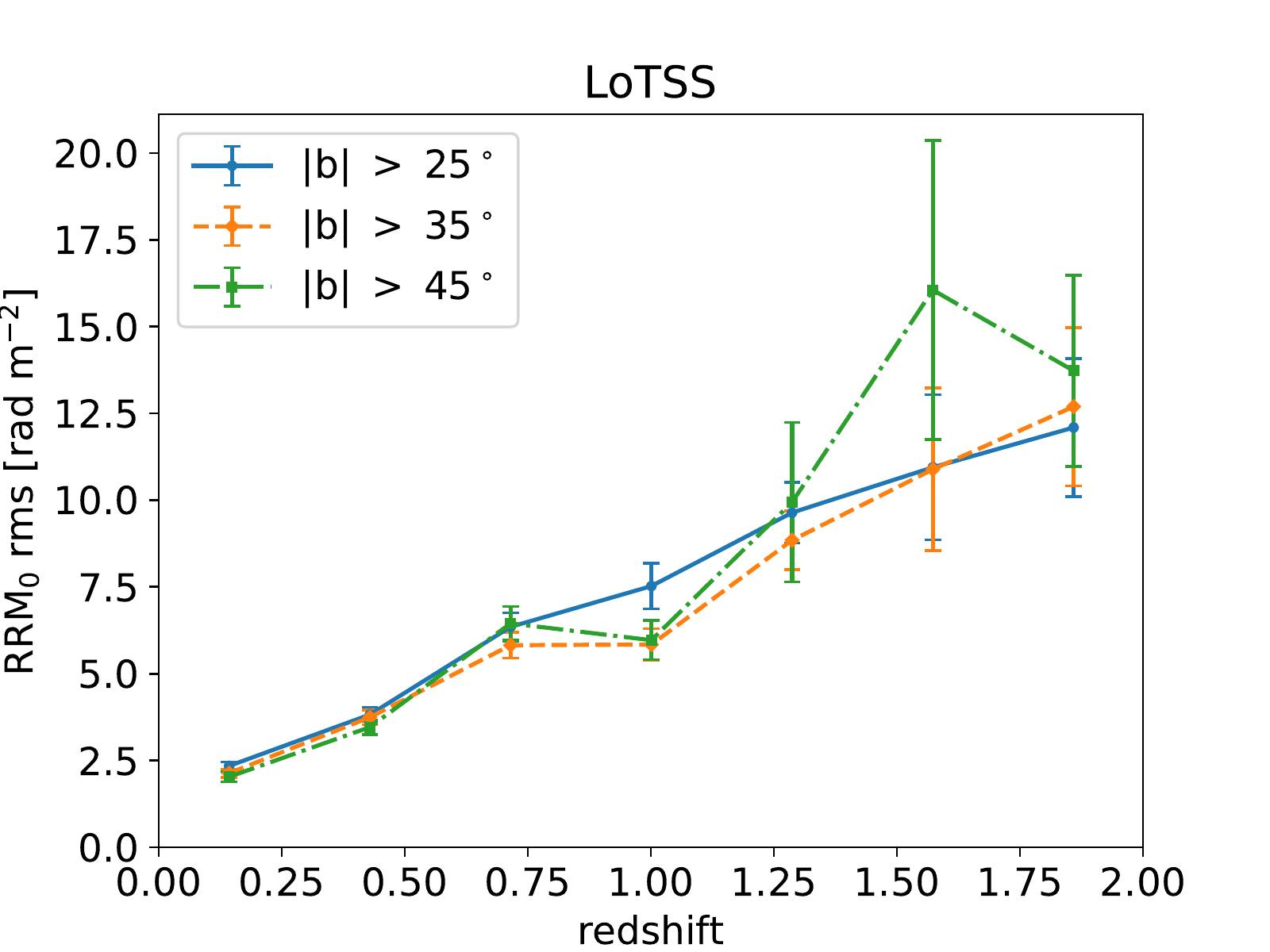}
   \caption{RRM$_0$ rms corrected for $C_1$ excluding sources at three different Galactic latitude ($b$) cuts.}
              \label{Fig:zdisp_gal}%
    \end{figure}
\begin{table*}
	\centering
	\caption{Linear best fit parameters of the rest frame RRM$_0$ rms  of the LoTSS sample for the three models $C_x, x=1,2,3$. The fit function is ${\rm RRM}_0 = \alpha + \beta\,z$, with $\alpha$ the intercept and $\beta$ the slope. The other parameters of the Table are the ratio $t$ between $\beta$ and its error $\sigma_\beta$, the Student's t-test probability that $t=0$ -- i.e. $\beta$ is flat ($p_t$), the Spearman's rank correlation coefficient $\rho$ between RRM$_0$ rms and  $z$, and its probability of uncorrelation ($p_\rho$). } 
	\label{tab:zfit}
	\begin{tabular}{lcccccc}
	  \hline 
        model & $\alpha $ &  $\beta$  &  $t = \beta/\sigma_\beta$  &  $p_t$  & $\rho$ & $p_\rho$ \\ 
           & [rad m$^{-2}$]  & [rad m$^{-2}$]  & & & & \\ 
       \hline 
        $C_1$ &  $1.68 \pm 0.31$  &  $5.84 \pm 0.27$  & 22.0  & $1.8\,\,10^{-6}$  & 1.00  &  $0.0$ \\ 
        $C_2$ &  $1.96 \pm 0.25$  &  $2.34 \pm 0.22$  & 10.7  & $6.2\,\,10^{-5}$  & 1.00  &  $0.0$ \\ 
        $C_3$ &  $2.23 \pm 0.31$  &  $1.33 \pm 0.27$  & 4.9  & $2.2\,\,10^{-3}$  & 0.96  &  $4.5\,\,10^{-4}$  \\ 
       \hline 
	\end{tabular}
\end{table*}

The RRMs must be corrected for redshift effects to get the  rest frame RRM$_0$ (seee Equation (\ref{eq:rm}) ). Specifically
\begin{equation}
    {\rm RRM_0=  RRM} (1+z_i)^2
\end{equation}
where $z_i <= z$ is the redshift at which the Faraday Rotation occurs along the line of sight and $z$ the source redshift. 
If the RRM is generated at more locations along the line of sight, then 
\begin{equation}
    {\rm RRM}_0 = \sum_i \Delta {\rm RRM}_i\, (1+z_i)^2
\end{equation}
where $\Delta$RRM$_i$ is the RRM contribution at redshift $z_i$.
What those redshifts are depends on how the medium is organised along the line of sight.  RRM$_0$ can be written as
\begin{equation}
   {\rm RRM}_0= C_x\,\, \rm RRM
\end{equation}
where $C_x$ is a correction factor depending on a model $x$ of the medium distribution. 

Three example cases bracket most of the possible models.
\begin{enumerate}
\item
In the simplest case the RRM occurs close to the source, either internal to the source or in the surrounding medium. This gives
\begin{equation}
    C_1 = (1+z)^2 , 
\end{equation}
where $z$ is the source redshift. 
\item 
At the other end, we  have the case in which the non-redshift-corrected  RRM is generated evenly along the line of sight,  which is the IGM in the foreground. This gives 
\begin{eqnarray}
    C_2 &=& \frac{1}{z}\int_0^z (1+z')^2\,dz'\\ 
        &=& \frac{(1+z)^3-1}{3z}
\end{eqnarray}
\item 
A third case is similar to (ii), except it is the rest frame RRM$_0$ that is evenly distributed.  More specifically, the two models  differ on how  RRM is distributed along the sight line. This model also describes the IGM and includes the scenario in which the RRM is generated by many cosmic web filaments along the line of sight. It gives
\begin{equation}
    C_3 = 1+z . 
\end{equation}
\end{enumerate}

Figure \ref{Fig:zdisp20}, bottom panel, and Table \ref{tab:zdisp} report the RRM rms in redshift bins corrected for $C_1$, $C_2$, and $C_3$. Data were first corrected and then binned. To check whether the increase with redshift is significant we fit them with a linear model, with the results shown in Table \ref{tab:zfit}.  The slope $\beta$ is non-flat  at high significance for all three cases ($\beta/\sigma_\beta  = 22.0$, 10.7, and 4.9 respectively, where $\sigma_\beta$ is the error of $\beta$) with Student's t-test probabilities of $p_t=1.8\times10^{-6}$,  $p_t=6.2\times10^{-5}$, and  $p_t=2.2\times10^{-3}$. Accounting for the redshift correction, there is  evolution of RRM$_0$ with redshift at high significance in the range $0 < z  < 2$ for all models we have considered. Spearman tests were also conducted with consistent results (see Table \ref{tab:zfit}). 

The analysis was repeated excluding the sources close to the  Galactic Plane for different Galactic  latitude ($b$) cuts. Figure \ref{Fig:zdisp_gal} shows the case of  RRM$_0$ with correction $C_1$. The behaviours are   similar and  the increase with redshift $z$ is present at all cuts. This excludes significant systematics from residual Galactic RM contamination, especially from the sources  at lower $|b|$. 

We computed the RRM mean and rms deviation also in redshift bins with equal-number of sources, with consistent results (see Appendix  \ref{app:RM_iso}).

\begin{figure}
   \centering
    \includegraphics[width=\columnwidth]{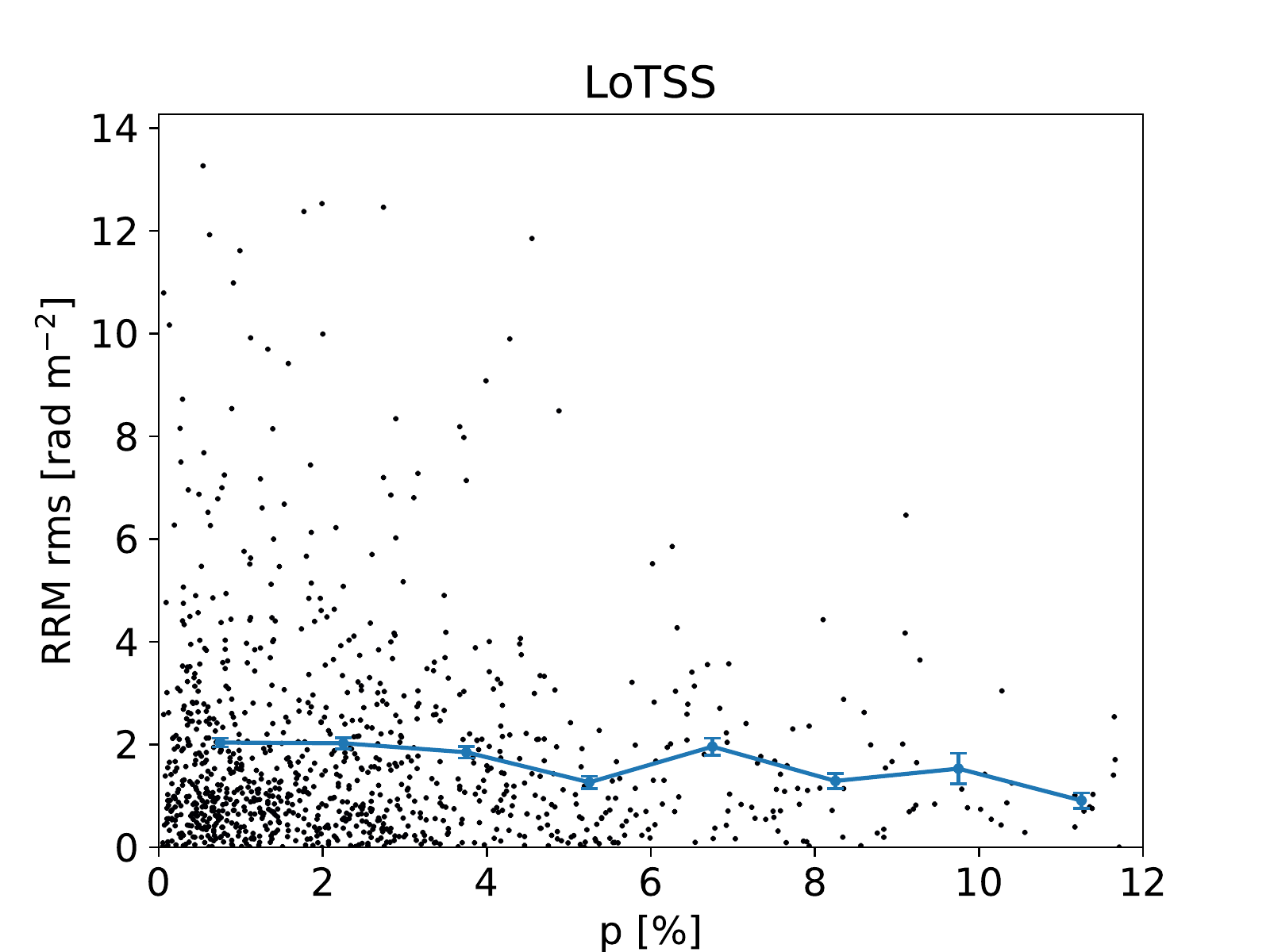}
    \includegraphics[width=\columnwidth]{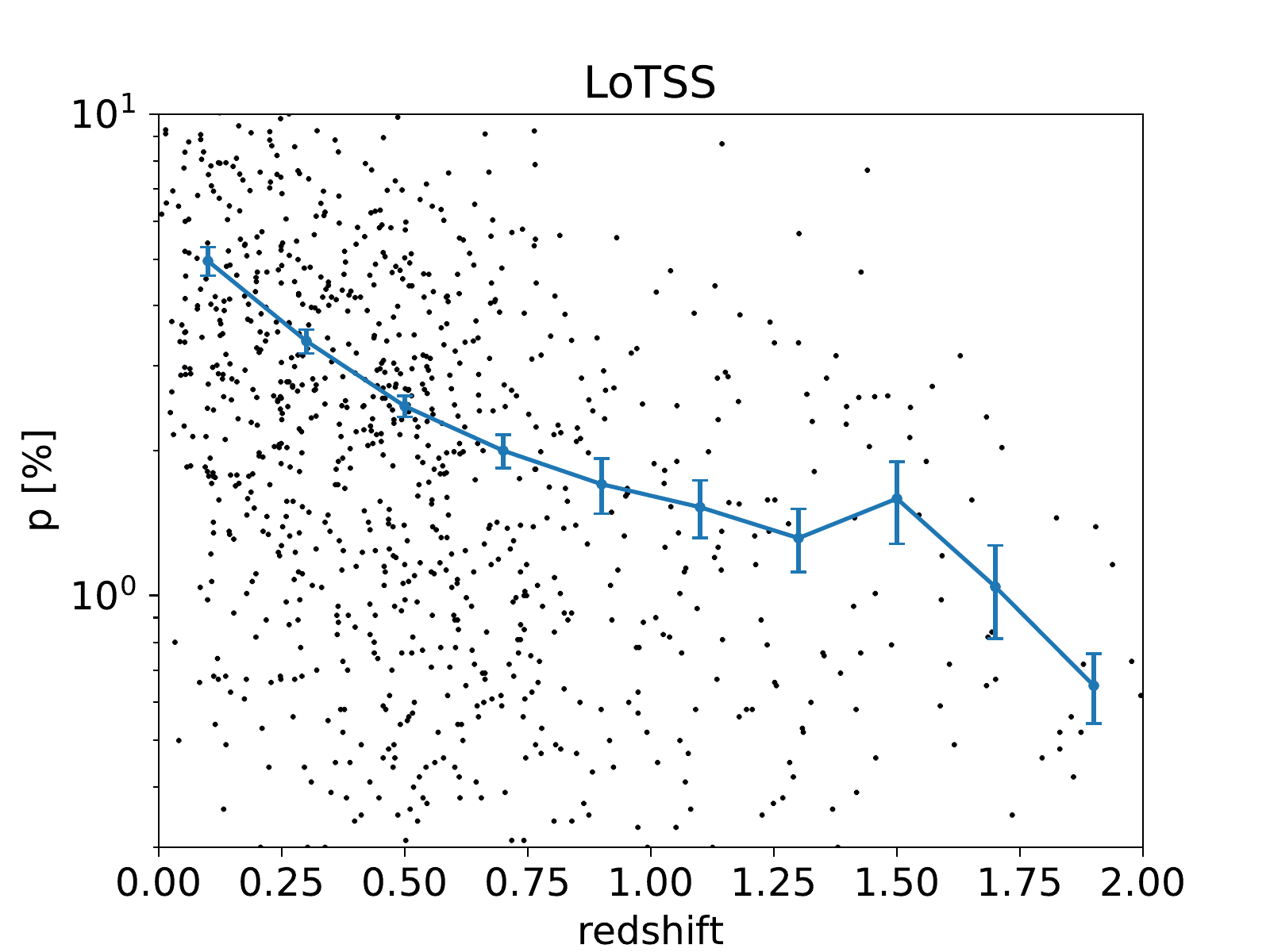}
   \caption{{\em Top}: RRM rms deviation as a function of fractional polarization $p$ of the LoTSS sample (solid line). Individual  $|$RRM$|$s  are also reported (dots). {\em Bottom}: Mean of the fractional polarization $p$ as a function of $z$ of the LoTSS sample  (solid line). Individual fractional polarizations are also reported (dots).}
              \label{Fig:pdisp}%
              \label{Fig:pmean}%
    \end{figure}

An analysis of the  fractional polarization evolution can help us with the interpretation, because a change with redshift would mean a change of the physical conditions at the sources or of the IGM over cosmic time. 
Figure \ref{Fig:pdisp}, top panel, shows $ \left< {\rm RRM}^2 \right>^{1/2}$ as a function of the fractional polarization $p$, with no obvious correlation.
The mean of the fractional polarization  as a  function of redshift is also shown  in Figure \ref{Fig:pmean}, bottom panel. There is clear anti-correlation, $p$ is highest at low $z$ (some 5\% at $z = 0.1$) and steadily drops towards high redshift (0.65\% at $z=1.9$). 
The behaviour is close to linear in log($p$)-$z$  space, the slope of the linear regression is $\beta = -0.396\pm0.044$ with ratio $\beta/\sigma_\beta = 9.0$ and t-test $p_t=8.9\times10^{-6}$. The Spearman's rank is $\rho = -0.96$ and $p_\rho = 7.3\times10^{-6}$. The decrease  is thus detected  at a high confidence level. 

\begin{figure}
   \centering
    \includegraphics[width=\columnwidth]{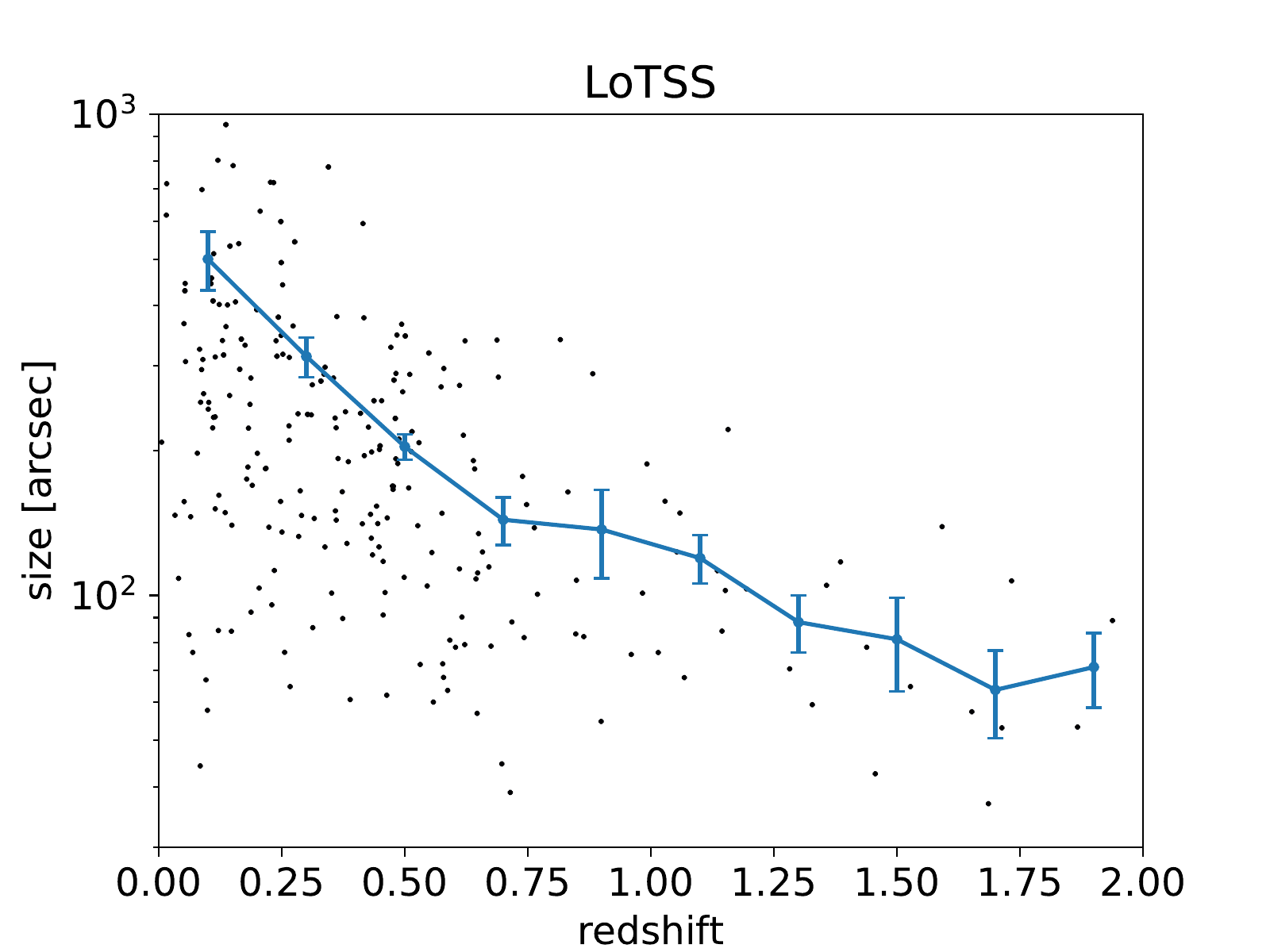}
   \caption{Mean source angular size as a function of $z$ of the LoTSS sample (solid line). Individual sizes are also reported (dots).}
              \label{Fig:smean}%
    \end{figure}

The mean source angular size in redshift bins is reported in Figure \ref{Fig:smean}. Only 269 sources have size information available. A decrement with $z$ is obvious from $\approx 500$~arcsec at $z=0.1$ to $\approx 70$~arcsec at $z=1.9$, making a correlation with depolarization possible. 
We computed also the physical sizes. The sources are very extended on average with mean sizes that run from $\sim 1.3$-Mpc at low redshift to $\sim 0.7$-Mpc at high redshift. It is clear that the objects for which the size information is available  (269 out of 1003, 27\%) are only extended sources and with the compact sources absent from this sub-sample.

We separated the sources into blazars and radio galaxies (i.e.~all other radio sources such as radio galaxies, Seyferts, QSOs, AGN) to broadly divide them in compact and extended sources. We used the classification from the BZCAT blazar catalogue\footnote{https://www.ssdc.asi.it/bzcat/}, SIMBAD database\footnote{http://simbad.u-strasbg.fr/simbad/}, and Sloan Digital Sky Survey (SDSS) DR16 catalogue\footnote{https://www.sdss.org/dr16/} -- with this priority respectively -- included in the LoTSS RM catalogue. All sources of our sample have a classification, of which 17\%  are blazars. Radio galaxies dominate except at $z\gtrsim 1.5$ where the number of blazars becomes comparable.

   \begin{figure}
   \centering
    \includegraphics[width=\columnwidth]{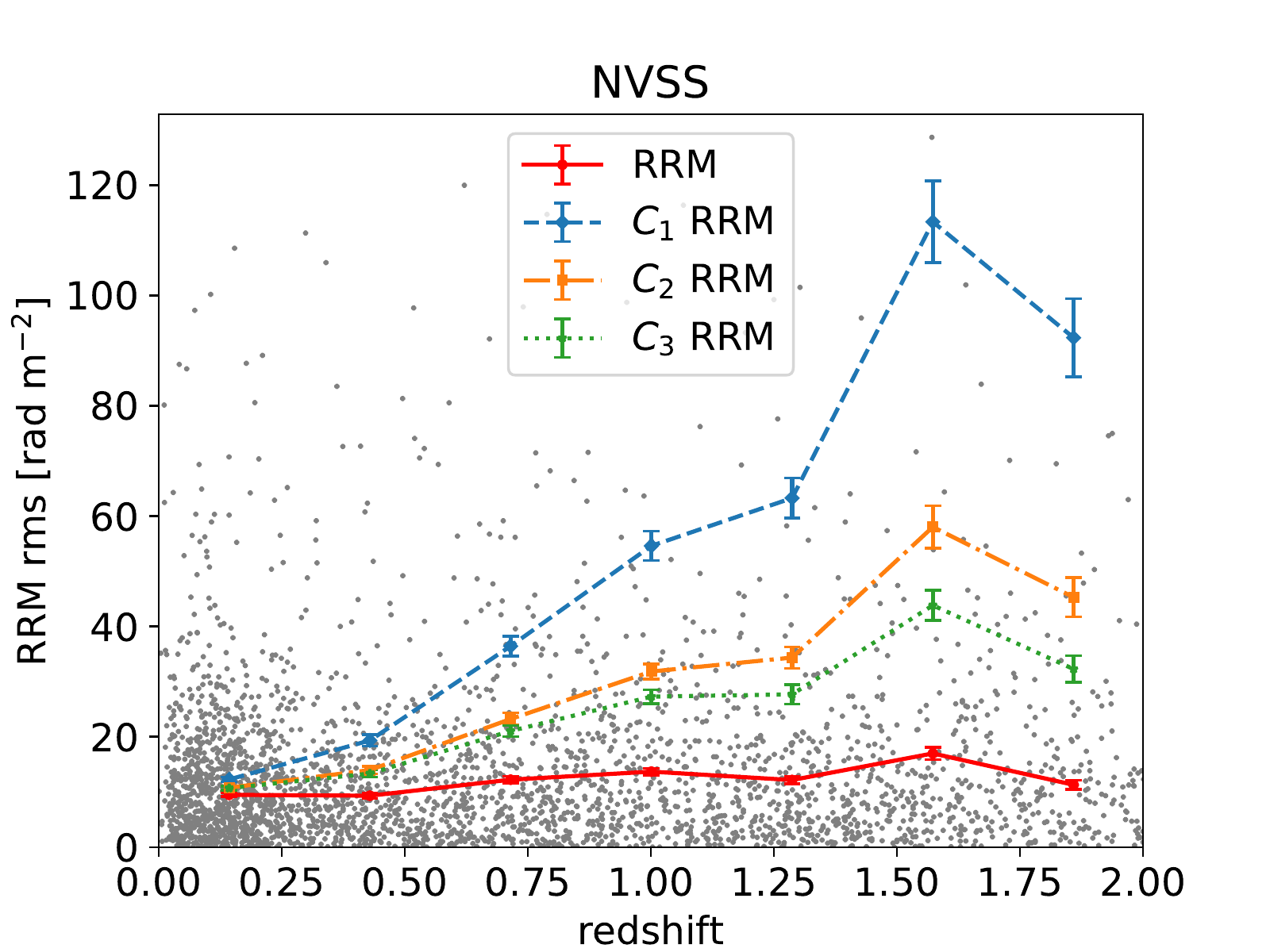}
    \includegraphics[width=\columnwidth]{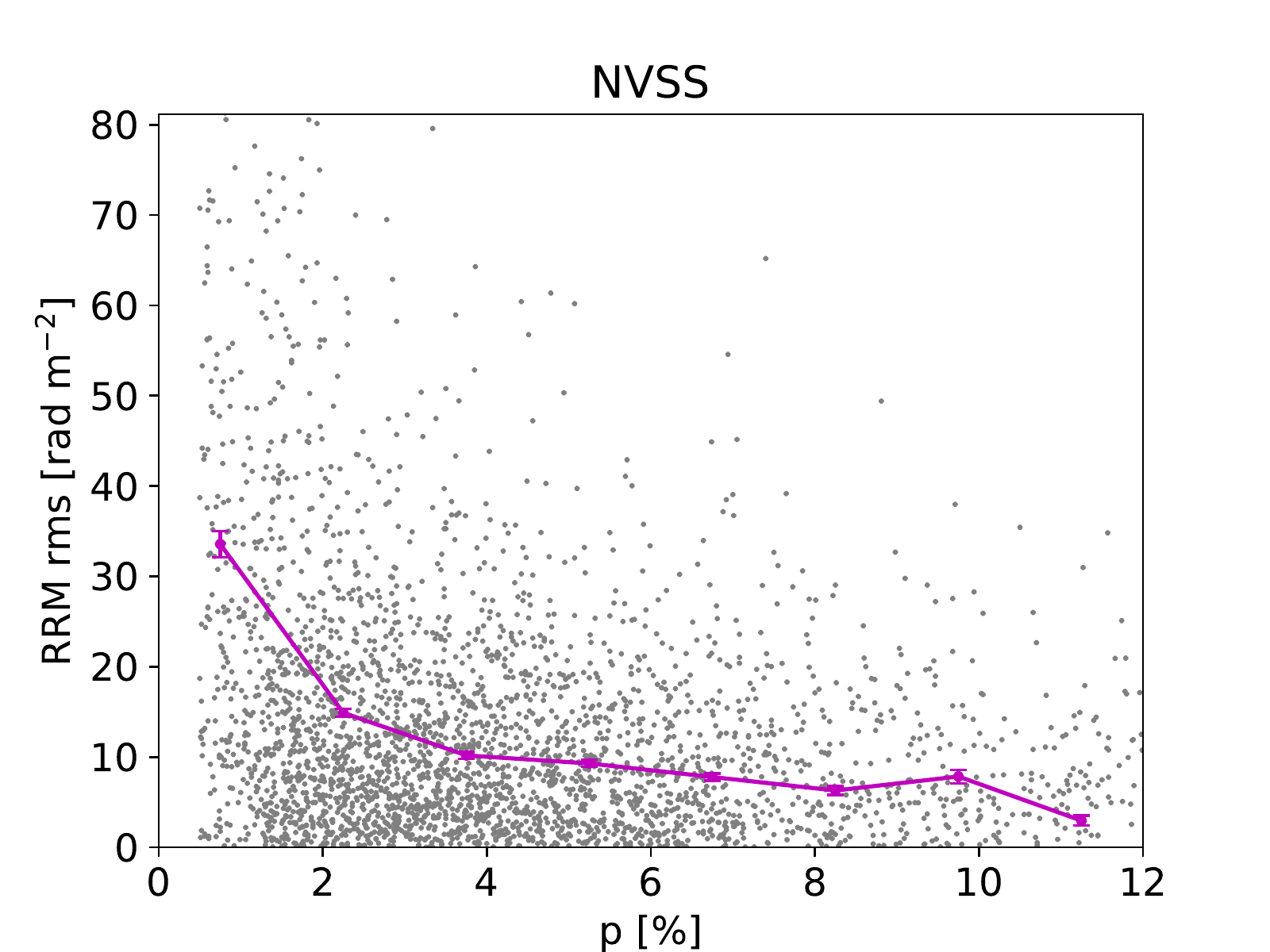}
    \includegraphics[width=\columnwidth]{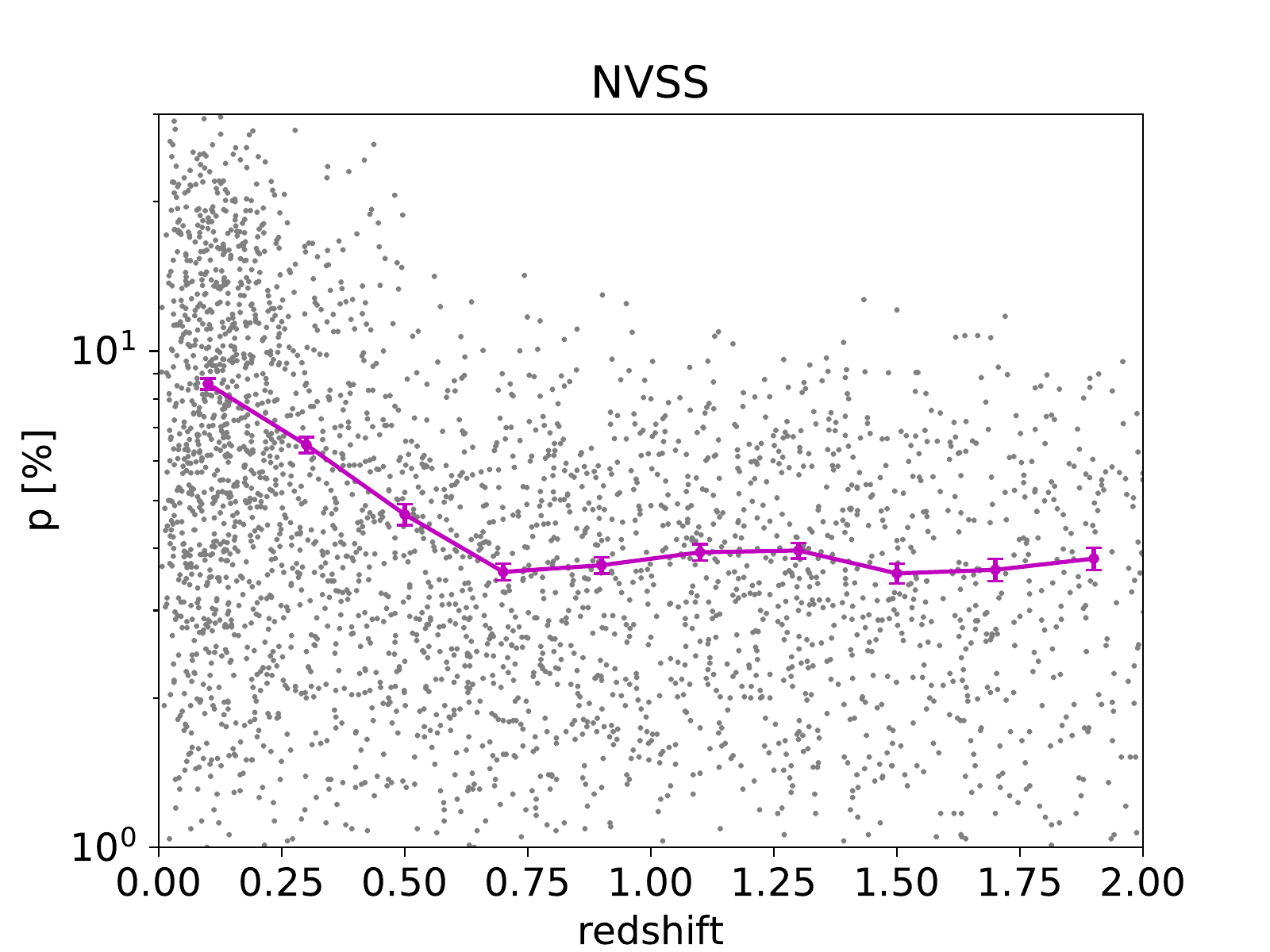}
   \caption{{\em Top}: NVSS sample RRM rms  deviation in redshift bins (red solid line).  The rest frame RRM$_0$ corrected for the redshift effect $C_1$ (dashed line),  $C_2$ (dot-dashed line), and $C_3$ (dotted) are also plotted. The individual |RRM|s  are reported as dots. {\em Middle}: RRM rms deviation as a function of fractional polarization $p$ for the NVSS sample (solid line). Individual  $|$RRM$|$  values are also reported (dots). {\em Bottom}: Mean of fractional polarization $p$ as a function of $z$ for the NVSS sample (solid line). Individual fractional polarization values are also shown (dots).}
              \label{Fig:zdisp_nvss_all}%
              \label{Fig:pdisp_nvss_all}%
              \label{Fig:pmean_nvss_all}%
    \end{figure}
   \begin{figure}
   \centering
    \includegraphics[width=\columnwidth]{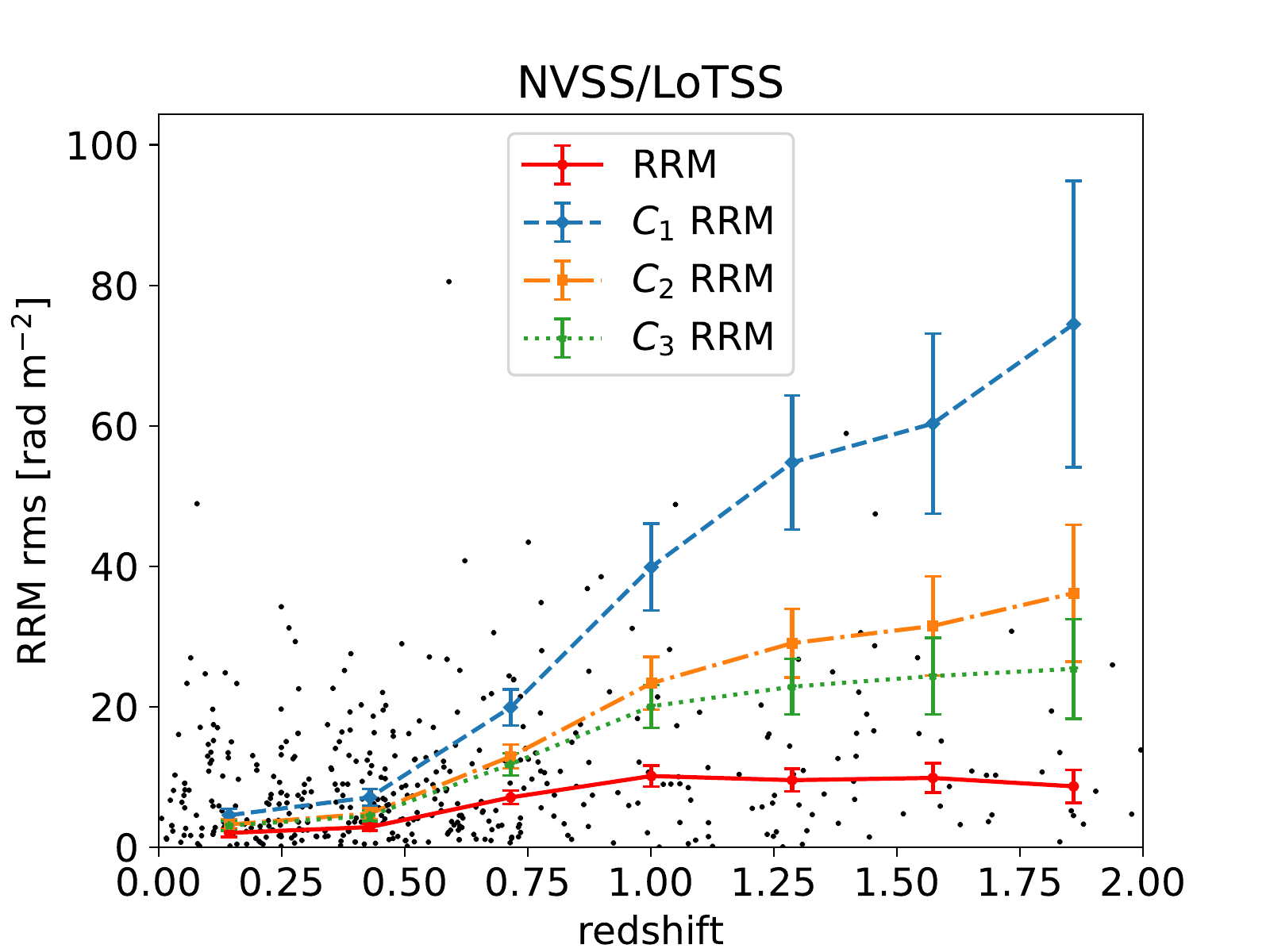}
    \includegraphics[width=\columnwidth]{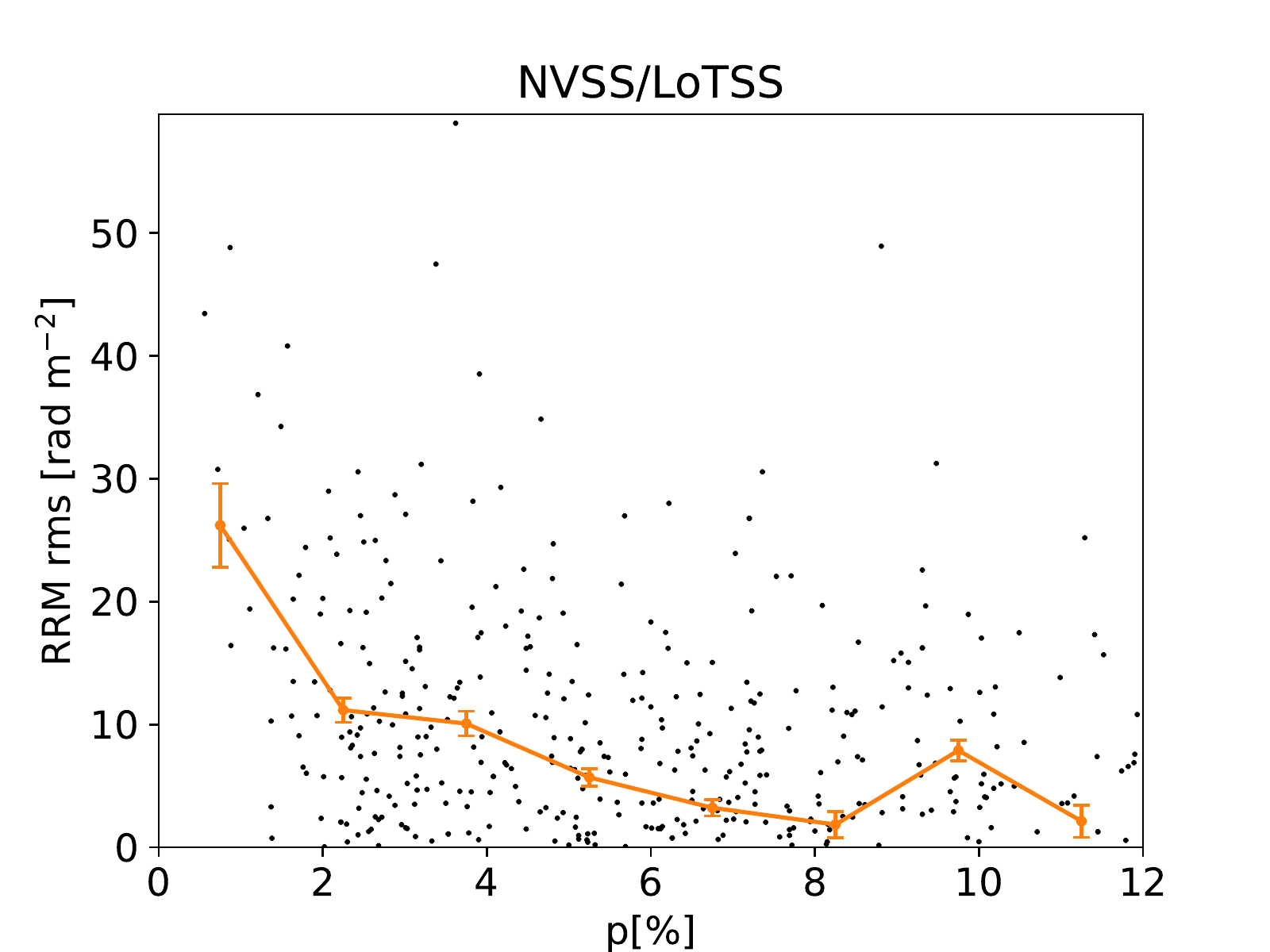}
    \includegraphics[width=\columnwidth]{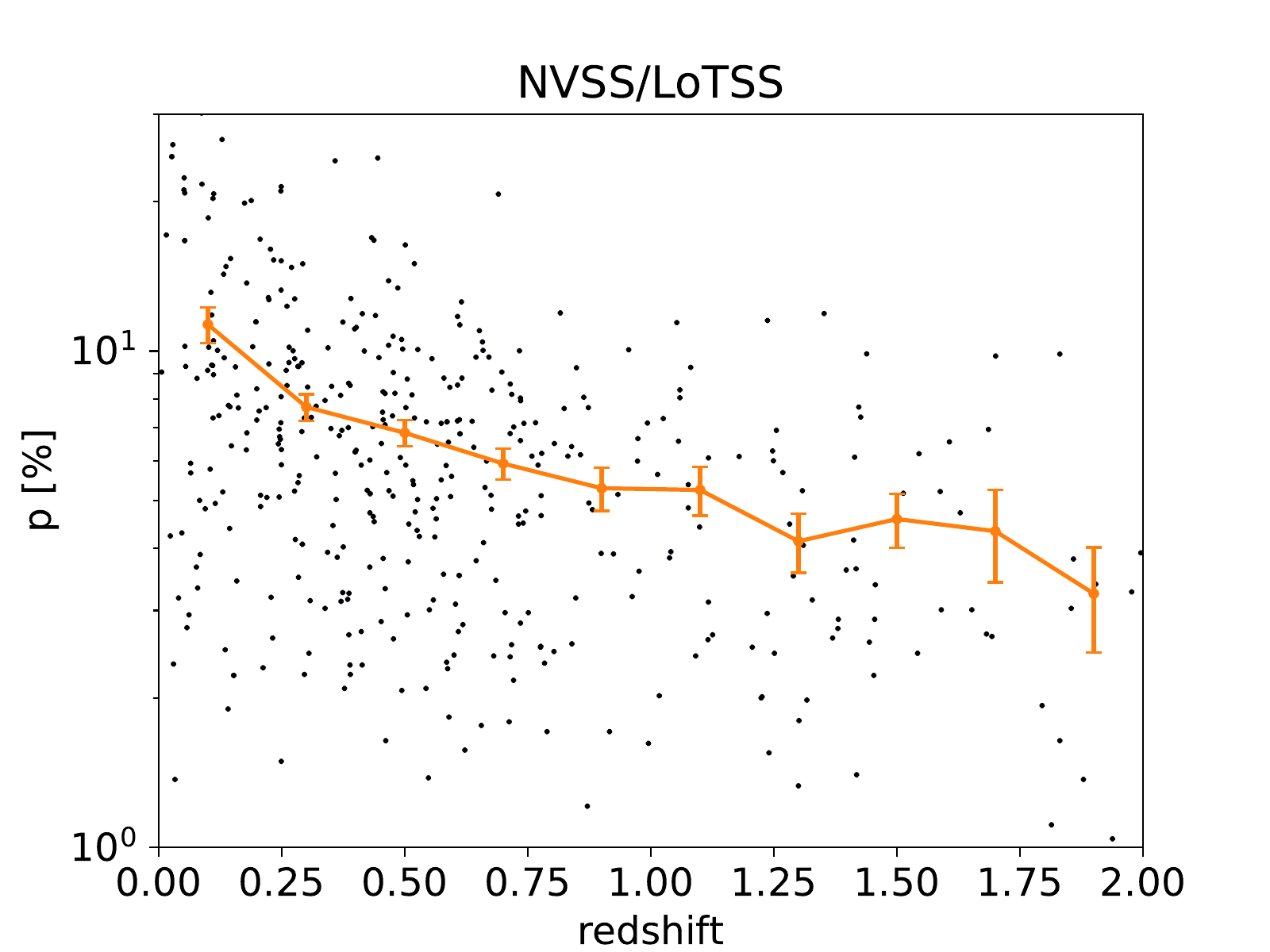}
   \caption{{\em Top}: NVSS/LoTSS sample RRM rms deviation as a function of redshift  (red solid line). The rest frame RRM$_0$ corrected for the redshift effect $C_1$ (dashed line), $C_2$ (dot-dashed line), and $C_3$ (dotted line) are also plotted. Individual |RRM| values  are reported as dots. 
   {\em Middle}: RRM rms deviation as a function of fractional polarization $p$ for the NVSS/LoTSS sample (solid line). Individual  |RRM|  are also reported (dots).
   {\em Bottom}: Mean of fractional polarization $p$ as a function of $z$ for the NVSS/LoTSS sample (solid line). Individual fractional polarization are also reported (dots).}
              \label{Fig:zdisp_nvss}%
              \label{Fig:pdisp_nvss}%
              \label{Fig:pmean_nvss}%
    \end{figure}

\subsection{NVSS}
The RRMs of the NVSS RM--redshift catalogue \citep{2012arXiv1209.1438H} were  computed estimating the GRM as in Section \ref{sec:rmvz_LoTSS}. 

The RRM dispersion of the entire sample, corrected for the RM measurement error and the GRM$_1$ error, is
\begin{equation}
      \left< {\rm RRM^2_{NVSS_{all}}} \right>^{1/2} = 13.28\pm0.27\,\, \rm rad \, m^{-2}
\end{equation}
This is much larger than for the LoTSS sample, possibly because of the smaller  depolarization at  higher frequency and hence  the source population  comes from a more diverse environment. At  higher frequencies, the polarized emission can survive after passing through higher density environments and hence develop higher RMs \citep{2020A&A...638A..48S, 2020MNRAS.495.2607O}. 

Figure \ref{Fig:zdisp_nvss_all}, top panel, shows the RRM rms deviation in  redshift bins. The behavior is flat before correcting for redshift effects, while after correction with models $C_1$, $C_2$, and $C_3$ an evolution with $z$ is obvious, albeit  with smaller confidence compared to the LoTSS sample. Both linear fit and Spearman rank results are reported in Table \ref{tab:zfit_nvss_all}.

The middle panel of Figure \ref{Fig:pdisp_nvss_all} shows RRM rms versus  fractional polarization $p$. There is a clear anti-correlation, a decrease of RRM, initially steeper and  then shallower.
This behaviour is similar to the result of \citet{2012arXiv1209.1438H}, who found an anti-correlation up to the same $p$ value and then a flattening close to their noise floor that they did not subtract in their plots.  They associated that anti-correlation to depolarization:  higher RRMs means the polarized radiation goes through higher density and higher magnetic field environments, which gives higher depolarization. 

The fractional polarization   $p$ versus redshift (Figure \ref{Fig:pmean_nvss_all}, bottom panel) decreases up to $z\sim 0.7$, then it is mostly flat (as found by \citealt{2012arXiv1209.1438H}). This differs from the LoTSS sample where $p$ decreases with $z$ for the entire range.

\subsection{LoTSS -- NVSS overlap sample}
To better compare the NVSS and LoTSS samples, we have selected the LoTSS RM catalogue sources   with an NVSS RM entry and a spectroscopic redshift, and then analysed the NVSS RMs at 1.4-GHz. This NVSS sub-sample is thus restricted to the LoTSS sample and we call it the NVSS/LoTSS sample. It  consists of 437 sources, 427 of which have $z<2$. 

\begin{table*}
	\centering
	\caption{Linear best fit parameters of the rest frame RRM$_0$ rms of the NVSS sample for the three models $C_x, x=1,2,3$. The fit function is ${\rm RRM}_0 = \alpha + \beta\,z$, with $\alpha$ the intercept and $\beta$ the slope. The other parameters of the Table are the ratio between $\beta$ and its error $\sigma_\beta$ ($t$), the Student's t-test probability that $t=0$, i.e.~that $\beta$ is flat ($p_t$), the Spearman's rank correlation coefficient $\rho$ between RRM$_0$ and  $z$, and its probability of no correlation ($p_\rho$). } 
	\label{tab:zfit_nvss_all}
	\begin{tabular}{lcccccc}
        \hline
        model & $\alpha $ &  $\beta$  &  $t = \beta/\sigma_\beta$  &  $p_t$  & $\rho$ & $p_\rho$ \\ 
           & [rad m$^{-2}$]  & [rad m$^{-2}$]  & & & & \\ 
       \hline 
        $C_1$ &  $-0.91 \pm 10.26$  &  $56.80 \pm 8.90$  & 6.4  & $7.0\,\,10^{-4}$  & 0.96  &  $4.5\,\,10^{-4}$ \\ 
        $C_2$ &  $5.74 \pm 5.25$  &  $25.32 \pm 4.55$  & 5.6  & $1.3\,\,10^{-3}$  & 0.96  &  $4.5\,\,10^{-4}$  \\ 
        $C_3$ &  $8.64 \pm 4.17$  &  $16.53 \pm 3.62$  & 4.6  & $3.0\,\,10^{-3}$  & 0.96  &  $4.5\,\,10^{-4}$  \\ 
       \hline 
	\end{tabular}
\end{table*}
\begin{table*}
	\centering
	\caption{Linear best fit parameters of the rest frame RRM$_0$ rms of the NVSS/LoTSS sample for the three models $C_x, x=1,2,3$. The fit function is ${\rm RRM}_0 = \alpha + \beta\,z$, with $\alpha$ the intercept and $\beta$ the slope. The other parameters of the Table are the ratio between $\beta$ and its error $\sigma_\beta$ ($t$), the Student's t-test probability that $t=0$, i.e.~that $\beta$ is flat ($p_t$), the Spearman's rank correlation coefficient $\rho$ between RRM$_0$ and  $z$, and its probability of no correlation ($p_\rho$). } 
	\label{tab:zfit_nvss}
	\begin{tabular}{lcccccc}
        \hline
        model & $\alpha $ &  $\beta$  &  $t = \beta/\sigma_\beta$  &  $p_t$  & $\rho$ & $p_\rho$ \\ 
           & [rad m$^{-2}$]  & [rad m$^{-2}$]  & & & & \\ 
       \hline 
        $C_1$ &  $-6.5 \pm 3.5$  &  $43.8 \pm 3.1$  & 14.3  & $1.5\,\,10^{-5}$  & 1.00  &  $0.00$ \\ 
        $C_2$ &  $-0.8 \pm 2.1$  &  $21.0 \pm 1.8$  & 11.7  & $3.9\,\,10^{-5}$  & 1.00  &  $0.00$  \\ 
        $C_3$ &  $1.3 \pm 2.3$  &  $14.7 \pm 2.0$  & 7.5  & $3.3\,\,10^{-4}$  & 1.00  &  $0.00$ \\ 
       \hline 
	\end{tabular}
\end{table*}

The RRM rms of this sample is 
\begin{equation}
      \left< {\rm RRM^2_{NVSS/LoTSS}} \right>^{1/2} = 5.72\pm0.36\,\, \rm rad\,m^{-2}
      \label{eq:rrm_nvlt}
\end{equation}
This is less than half the value derived for  the whole NVSS sample, confirming that the LoTSS low frequency catalogue selects for low density environments that generate lower RRMs. 

Figure \ref{Fig:zdisp_nvss} shows the RRM rms as a function of $z$ and $p$, and $p$ as a function of redshift. It is worthwhile to note some differences  of these 1.4-GHz RRMs that are in low density environments. The RRM rms has a gentle increase with $z$, which is unseen in both the LoTSS and the whole NVSS sample. The increase is marginally significant with a slope that differs from zero by $3.3\,\sigma$ and a $p$-value of $p_t = 1.1\times10^{-2}$. The redshift-corrected RRM$_0$ has a high significance detection of an evolution with $z$ for all models we considered (Table \ref{tab:zfit_nvss}). 

The RRM  anti-correlates with $p$, as in the NVSS case and differing from the LoTSS sample, up to $p \sim 8$\% and then it flattens. 
The fractional polarization  decreases with $z$ at high significance, as at 144-MHz albeit at a lower rate (a factor $\sim 3$ between the two redshift range ends instead of $\sim 8$), but different from the whole 1.4-GHz sample that shows an initial evolution only. All points except one follow the decreasing trend and that single point is at some  1-sigma  from the general trend. A linear--log space linear fit  gives a slope of $\beta =-0.242\pm 0.029$ with ratio $\beta/\sigma_\beta=8.3$ and $p$-value of $p_t =1.7\times10^{-5}$. Spearman's rank is $\rho=-0.96$, $p_\rho=7.3\times10^{-6}$.

\begin{figure}
   \centering
    \includegraphics[width=\columnwidth]{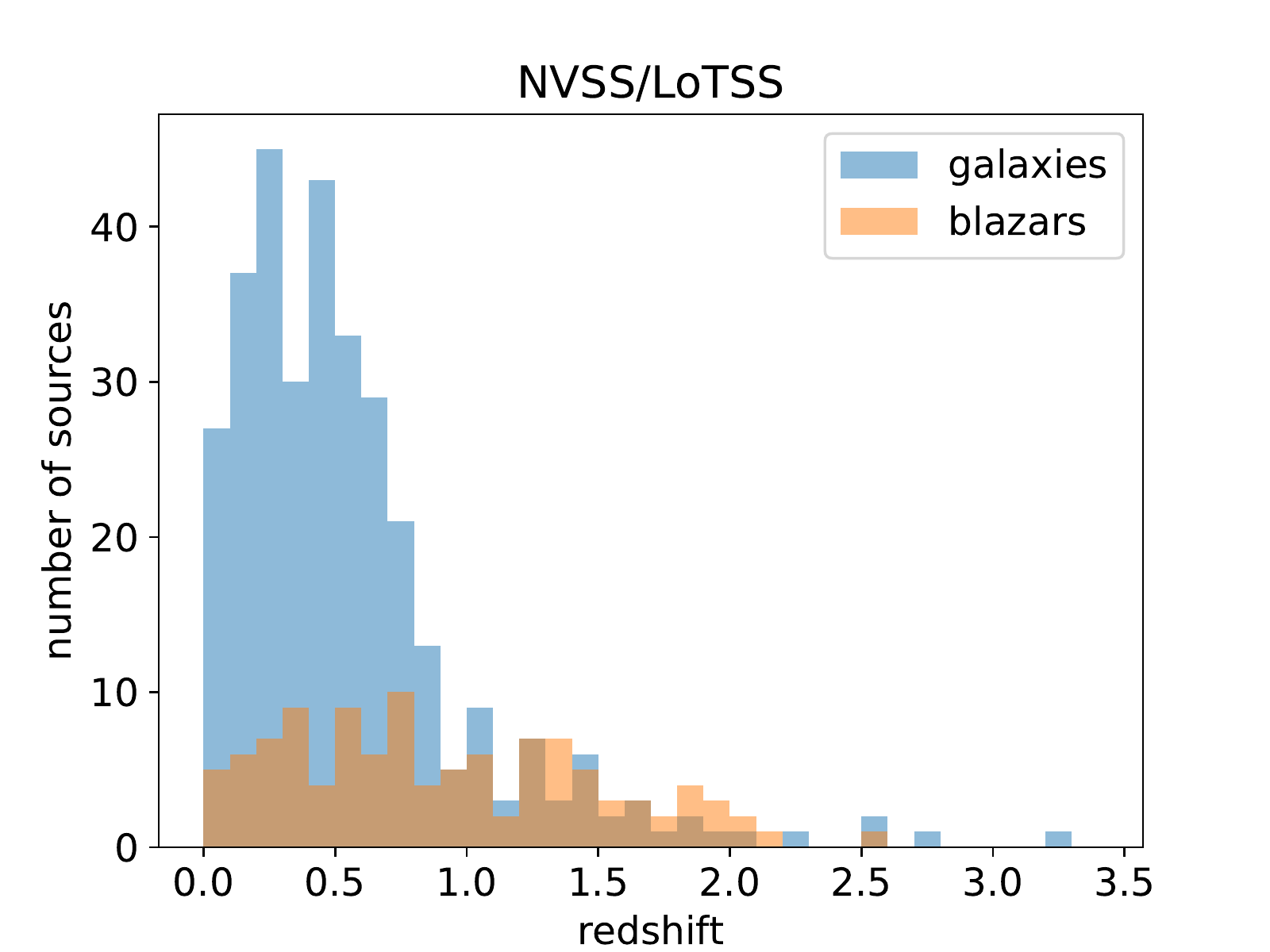}
   \caption{Redshift distribution of radio galaxies and blazars of the NVSS/LoTSS sample. }
              \label{Fig:zdistgbz_nvss}%
    \end{figure}

We also separated the sources into blazars and radio galaxies, as done for the LoTSS sample. All of the sources in the NVSS/LoTSS sample have classifications, with 25\% being blazars. The redshift distributions of the two groups are shown in Figure \ref{Fig:zdistgbz_nvss}, where radio galaxies dominate  out to  $z\approx 0.9$, above which blazars become comparable or dominant. 

   \begin{figure}
   \centering
    \includegraphics[width=\columnwidth]{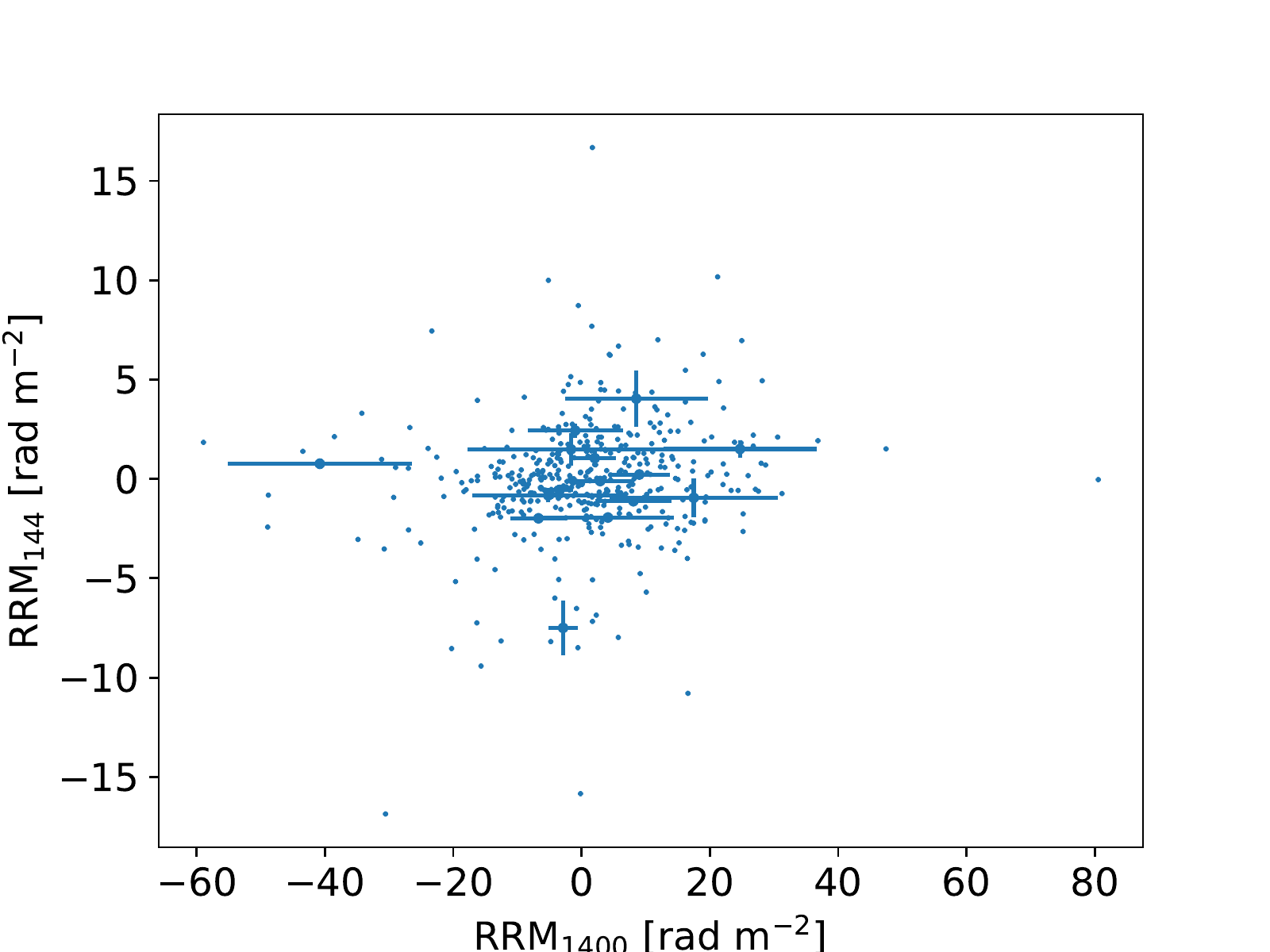}
   \caption{RRM at 144-MHz, RRM$_{144}$, plotted versus that at 1.4-GHz, RRM$_{1400}$, for the NVSS/LoTSS sample. Error bars are plotted 1-in-30 sources to avoid too much confusion.}
              \label{Fig:rrmplt}%
    \end{figure}

The large difference between the RRM rms at 144-MHz and 1.4-GHz (1.9 and 5.7-rad m$^{-2}$) suggests the RRMs at the two frequencies  are different. In Figure \ref{Fig:rrmplt} we plot them against each other. Indeed, the sources appear randomly distributed with no obvious trend, which would indicate the two sets of RRMs are different and thus of different origin. However, the large NVSS error bars cover most of the spread preventing us from drawing firm conclusions.

\section{Discussion}
\label{sec:discussion}

\subsection{Environment}\label{sec:disc_env}

The low RRM rms of $\sim$1.9-rad m$^{-2}$ that we measure for the LoTSS sample supports that the polarized emission at low frequency comes from and propagates through low density environments, where it  can survive depolarization, as found in earlier work \citep[e.g.][]{2020A&A...638A..48S,2020MNRAS.495.2607O}. This also appears to be supported by the higher frequency NVSS sample (1.4-GHz) that, once restricted to the sources in common with the LoTSS catalogue, has an  RRM rms  $\approx 2$ times smaller than the full sample. 

   \begin{figure}
   \centering
    \includegraphics[width=\columnwidth]{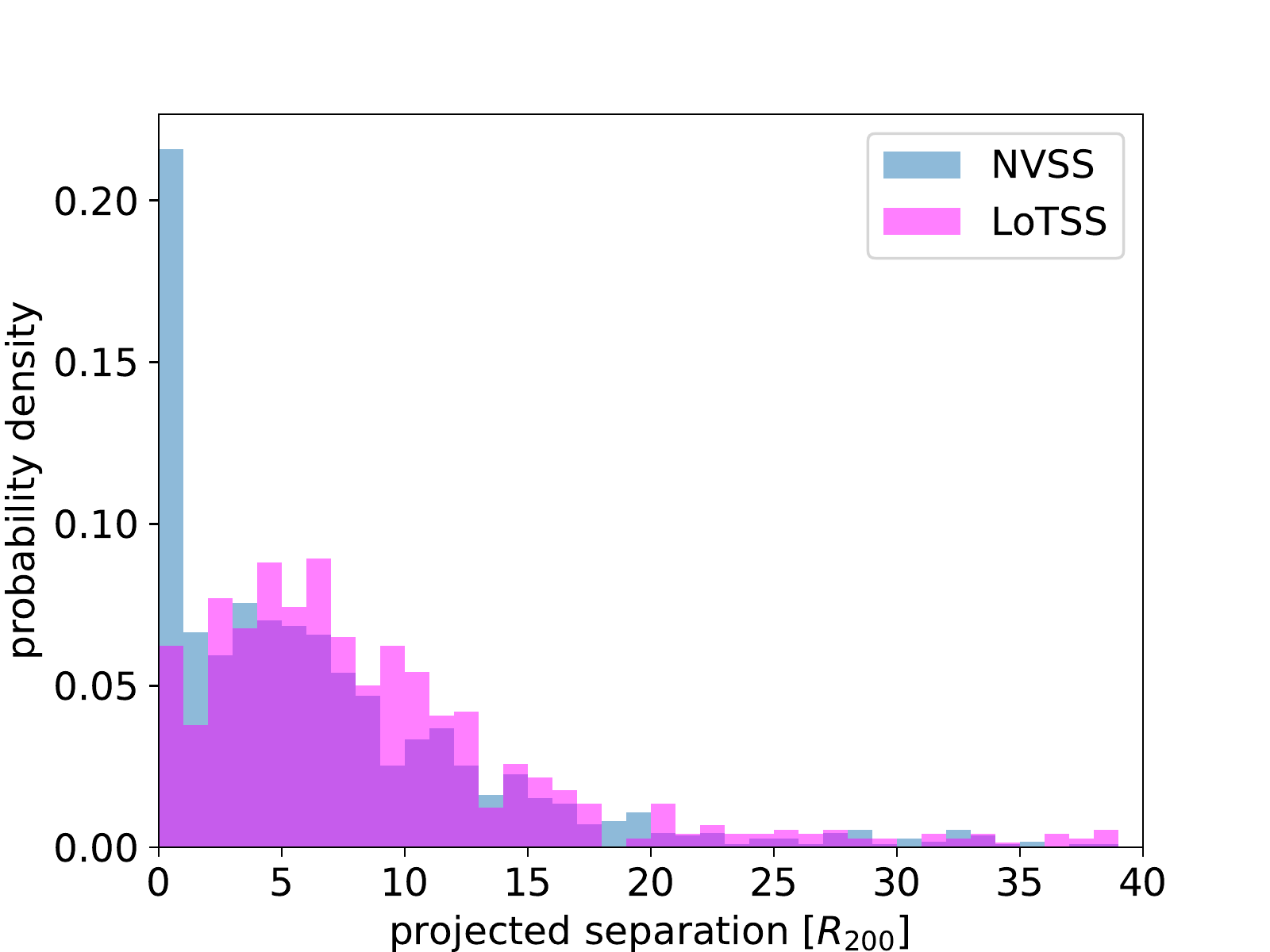}
   \caption{Distribution of the minimum projected separations of  sources of the LoTSS and NVSS samples from galaxy clusters. Separations are plotted up to $40 \,R_{200}$ for viewing reasons, there are 14 more sources of the LoTSS sample and 21 of the NVSS sample beyond that limit.}
              \label{Fig:rdist}%
    \end{figure}

To further support these considerations, we measured the projected separation in $R_{200}$\footnote{$R_{200}$ is the distance from the cluster centre where the density drops to 200 times the critical density of the Universe.} units for our sources from the nearest galaxy cluster, for both LoTSS and NVSS samples, where $R_{200}$ is approximately the outer boundary of galaxy clusters. We use  the catalogue of 158,103 clusters by \citet{2015ApJ...807..178W} \citep[see also][]{2012ApJS..199...34W} that spans  $0.05 < z <  0.75$ and  has a mix of spectroscopic and photometric redshifts with errors up to $ 0.018$.Masses of catalogue clusters are as low as $2\times 10^{12} M_\odot $ and the catalogue is more than 95\% complete for  masses larger than $1.0\times 10^{14} M_\odot $, which covers well down to poor clusters. Note that the catalogue gives $R_{500}$\footnote{$R_{500}$ is the distance from the galaxy cluster centre where the density is 500 times the critical density of the Universe.}  for each galaxy cluster and we estimate $R_{200}$ by  the relation $R_{500}/R_{200} \approx 0.7$ \citep{2009A&A...496..343E}. For each source  of redshift $z_{s}$ we search for the galaxy cluster with minimum projected separation  in the redshift range $|z_{gc}-z_s| < 0.036$ (2--sigma uncertainty), where   $z_{gc}$ is the cluster redshift. We restrict our search to sources with redshift in the catalogue range and that are in the catalogue footprint, resulting in  739 (LoTSS) and 1116  (NVSS) sources.  The minimum projected separation distributions are shown is Figure \ref{Fig:rdist}.  Only a small fraction of sources (6.2\%) in the LoTSS sample is within $R_{200}$ from the nearest cluster, which increases to 21.5\% for the NVSS sample. The median separation is 7.0  and 5.2~$R_{200}$ for LoTSS and NVSS, respectively. The distribution of the LoTSS sample peaks at $\sim 5\,R_{200}$ and then decreases towards separation 0, while for the NVSS sample it is mostly flat down to the smallest separations, meaning that the peak is closer to separation 0. Overall,  the 144-MHz LoTSS sources tend to reside far from galaxy clusters, in low density environments, while the 1.4-GHz NVSS sources are closer to clusters with a marked peak within cluster boundaries. 
We do not have available an equivalent catalogue of galaxy groups and we cannot conduct a similar analysis for them.

   \begin{figure*}
   \centering
    \includegraphics[width=\columnwidth]{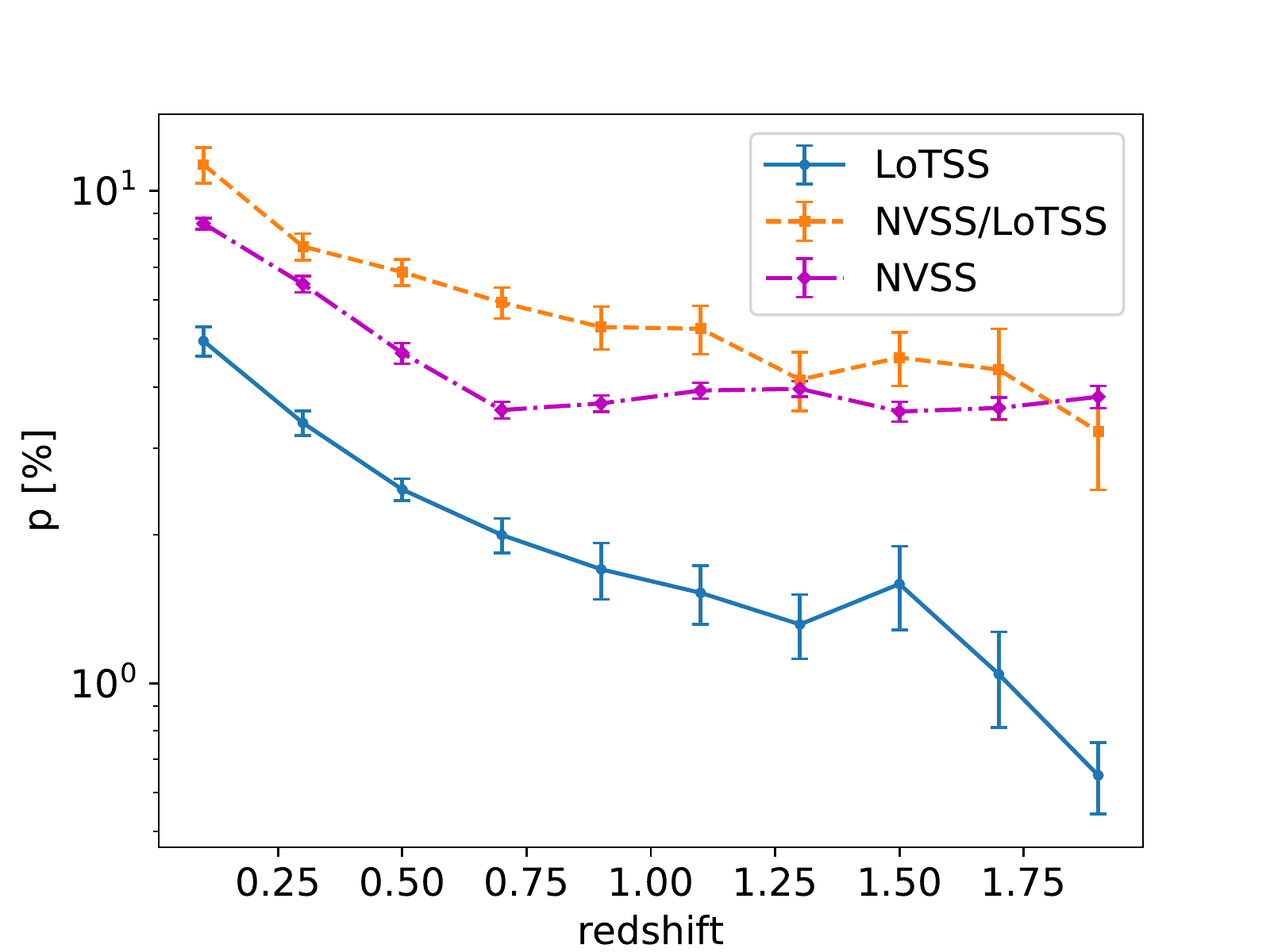}
    \includegraphics[width=\columnwidth]{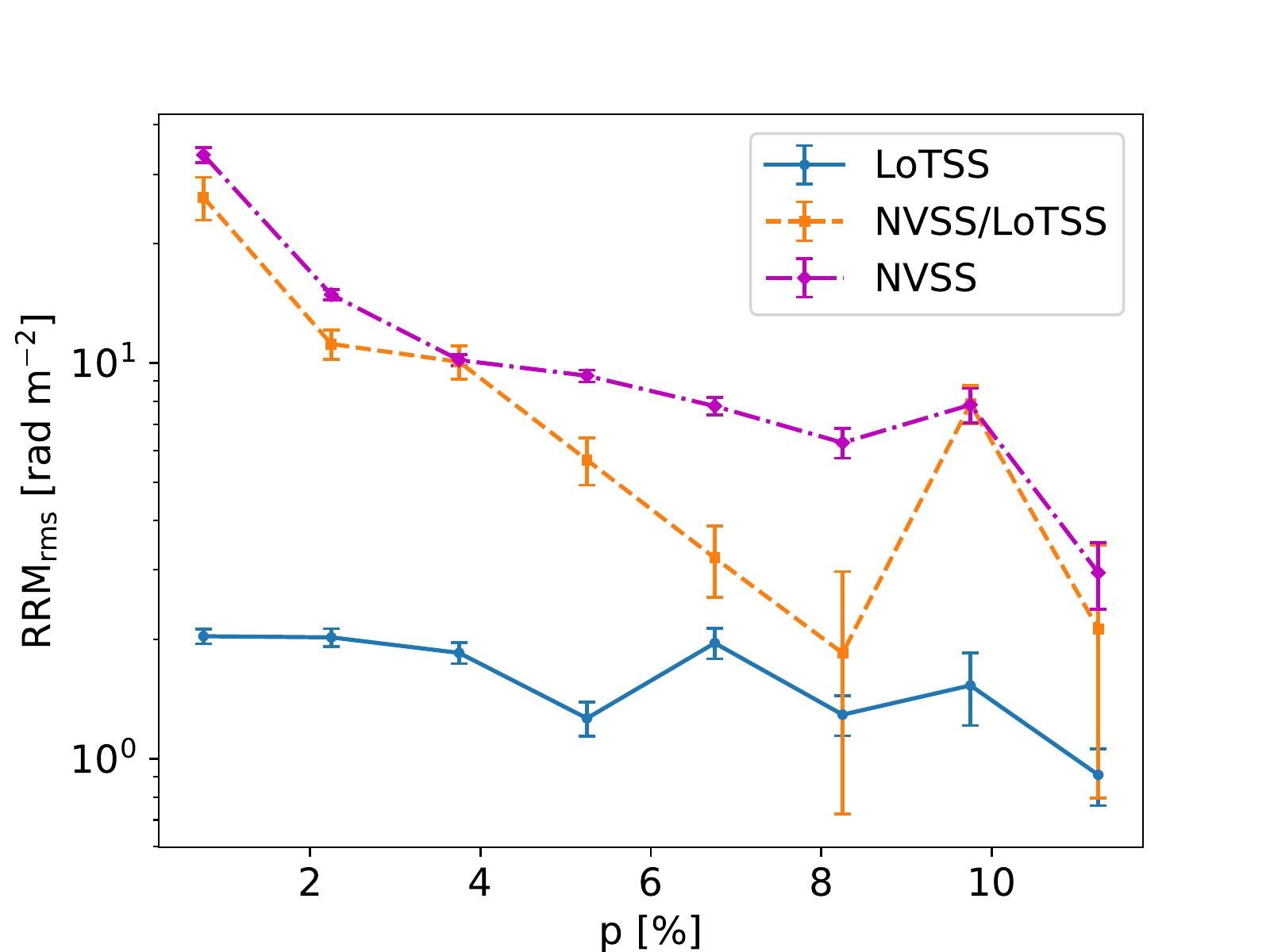}
   \caption{{\em Left}: Mean of fractional polarization $p$ as a function of $z$ for all the samples used in the paper (see legend).
   {\em Right}: RRM deviation (RRM$_{\rm rms}$) as a function of $p$ for all the samples used in the paper (see legend).}
              \label{Fig:pall}%
              \label{Fig:rrmpall}%
    \end{figure*}

Considerations based on simple depolarization models bear  similar conclusions. These are approximations of real cases, but are useful to give first order estimates. As mentioned previously,  the extragalactic RM can be generated locally to the source or by the IGM through which the radiation propagates.  In the former case, most of the RM and $p$ are generated  in the environment surrounding the source \citep{2008MNRAS.391..521L}. Both this and the radiation propagating through the  IGM  can be modeled  by a turbulent slab (external Faraday dispersion) whose  depolarization is described by the equation \citep{1966MNRAS.133...67B,1998MNRAS.299..189S}
\begin{equation}
    \frac{p}{p_0}=e^{-2\,\sigma_{\rm RM}^2\,\lambda^4}
    \label{eq:sigrm}
\end{equation}
where  $p$ and $p_0$  are the emerging polarization fraction and that of the radiation entering the region, respectively. The $\sigma_{\rm RM}$ parameter is the RM dispersion of the region, and $\lambda$ is the observing  wavelength. Assuming the signal is totally depolarized for\footnote{We assumed this value because a depolarization of 10\% looked insufficient and 1\% too much at this frequency where the typical fractional polarization is of a few percent.} $p/p_0 =1/30$, the signal can survive  depolarization at 144-MHz for $\sigma_{\rm RM}  < 0.3 \,{\rm rad\,m}^{-2}$, which sets a small limit for the RM turbulence it can go through and thus requires low density environments. 

\subsection{Fractional polarization behaviour}
\label{sec:disc_gc}
There is a clear evolution with redshift of $p$ and of the rest frame RRM$_0$ for the LoTSS sample, for any redshift correction model we use. This might happen local to the source or in the IGM along the line of sight. A comparison with the results at 1.4-GHz can help with the interpretation. 

Figure \ref{Fig:pall}, left panel, shows the behaviour of $p$ versus $z$ for all of our three samples. LoTSS has the lowest values   with a steady decrement with redshift,  while NVSS has higher values that, after an initial decrement,  becomes flat. NVSS/LoTSS  is always higher than NVSS, as expected since the emission is coming from lower density environments, and it shows a steady decrease with redshift, similar to LoTSS, albeit at a significantly lower rate. 

There are two possible explanations for the depolarization behaviour of the LoTSS sample, an astrophysical origin or beam  depolarization because the source size gets smaller at higher redshift (Figure \ref{Fig:smean}). The latter is a frequency independent effect. 
The decrease of $p(z)$ and its flattening at high redshift for the NVSS sample was interpreted by \citet{2012arXiv1209.1438H} as a mixing of two populations  with different polarization fractions: radio galaxies at low redshift and compact sources that have a lower polarization fraction  at high redshift. The NVSS/LoTSS sample has a continuous decrement with redshift. Radio galaxies are the dominant population up to $z\sim 0.9$ and beam depolarization could explain it, but at $z>0.9$  blazars are comparable or  dominate  and a flattening would be expected. Similarly, the LoTSS sample has a transition at $z\sim 1.5$,  but no flattening is observed.  In this context, we note that the angular diameter distance peaks at $z \sim 1.5$ and is quite flat in the range $z= [1,2]$,
thus  beam depolarization alone cannot explain the drop of $p$ in that range. Finally, for the  NVSS/LoTSS sample, that traces the similar environments of the full LoTSS sample, $p$ anti-correlates with redshift as for LoTSS, but at a smaller rate, thus at most only a minor component can be due to frequency independent depolarization. 
Note that the coarser beam at 1.4-GHz can  generate more depolarization and correcting for it would further increase the difference in the $p$ decrement rate between the two frequencies, reinforcing  the conclusion that only a minor component of the depolarization at 144-MHz can be attributed to beam depolarization. 
We conclude that beam depolarization is unlikely to be the cause of most of the behaviour with redshift for $p$, leaving the astrophysical origin as the most likely explanation.  
An anti-correlation at 1.4-GHz was also found by \citet{2021A&A...653A.155B} with a much smaller, lower flux sample and they too excluded beam depolarization. 

As mentioned, the astrophysical origin of the $p$--$z$ anti-correlation can be either  local to the source or in the IGM between us and the source. At 1.4-GHz, we can exclude the latter as the dominant term. The depolarization of the NVSS/LoTSS sample at $z=1.9$ is $p/p_0\sim30$\%, measured as the ratio of the mean of $p$ at the high and low redshift end. If this is due to the IGM, from  Equation (\ref{eq:sigrm}) the depolarization would drop to 0.003\% at 144-MHz and we would not see any polarized signal. Hence, the depolarization observed  at 1.4-GHz must be  local to the source and the components observed at 144 and 1400-MHz must be of a different nature.
The component that we see at 144-MHz has almost no depolarization at 1.4-GHz.
At 144-MHz, the depolarization is still compatible with either possible origin, and thus not inconsistent with being generated by the IGM. We expect the IGM to consist of filaments whose number increases with $z$ leading to increasing depolarization with $z$.  
Among local origin effects, a couple can be excluded. 
At both 144-MHz and 1400-MHz, the behaviour of $p$ is unlikely to be due to a change of population with redshift. As mentioned above, a flattening would be expected at high redshift, which is not observed. Also effects from external Faraday dispersion (Equation  (\ref{eq:sigrm})) are unlikely because at high redshifts the rest frame frequency is higher by a factor of $(1+z)^2$ and the depolarization is expected to be smaller, while the opposite is observed.

The behaviour of RRM versus $p$ is different at 144-MHz and 1400-MHz (Figure \ref{Fig:rrmpall}, right panel, shows all of the three samples). It is anti-correlated with $p$ at 1400-MHz for both the NVSS and NVSS/LoTSS sample.  That behaviour was associated by \citet{2012arXiv1209.1438H}  to the effect of depolarization:  higher RRMs means the polarized radiation goes through higher density and higher magnetic field environments.This usually gives higher turbulence and RRM dispersion, and in turn    higher depolarization (e.g., Equation (\ref{eq:sigrm})). This points to the RRM being  generated at the source at  1.4-GHz. At 144-MHz it is flat and the RRM is independent of depolarization. This is a totally different behaviour and again points to the RRM generation mechanism being different to that at higher frequency.
Cosmological MHD simulations find that the RRM generated by filaments of the cosmic web along the line of sight is mostly  independent of redshift \citep{2011ApJ...738..134A}, because the increase of rest frame RRM$_0$ with redshift 
(because of the higher number of filaments intercepted)
is compensated by the redshift correction. RRM is also expected to be independent of $p$ because RRM is uncorreleted with $z$ while $p$ changes. The flat behaviour of RRM versus $p$ and $z$ is thus consistent with that expected for a  IGM/filaments scenario at 144-MHz and against a local origin. Notice that for the 1.4-GHz NVSS/LoTSS sample the RRM marginally increases with redshift, inconsistent with a generation dominated  by the IGM.

In the next sections, we analyse in detail the two possible scenarios at 144-MHz that could explain the behaviour that we observe for $p$, RRM, and RRM$_0$.

\subsection{IGM filaments}
\label{sec:disc_cwf}

Several arguments, as  described above, point to the RRM and $p$ we observe at 144-MHz being consistent with a generation  from filaments of the cosmic web, e.g.~the flat behaviour of RRM with $p$ and $z$.  Assuming that, we can derive some properties of the magnetic field of filaments. 

\begin{figure}
   \centering
    \includegraphics[width=\columnwidth]{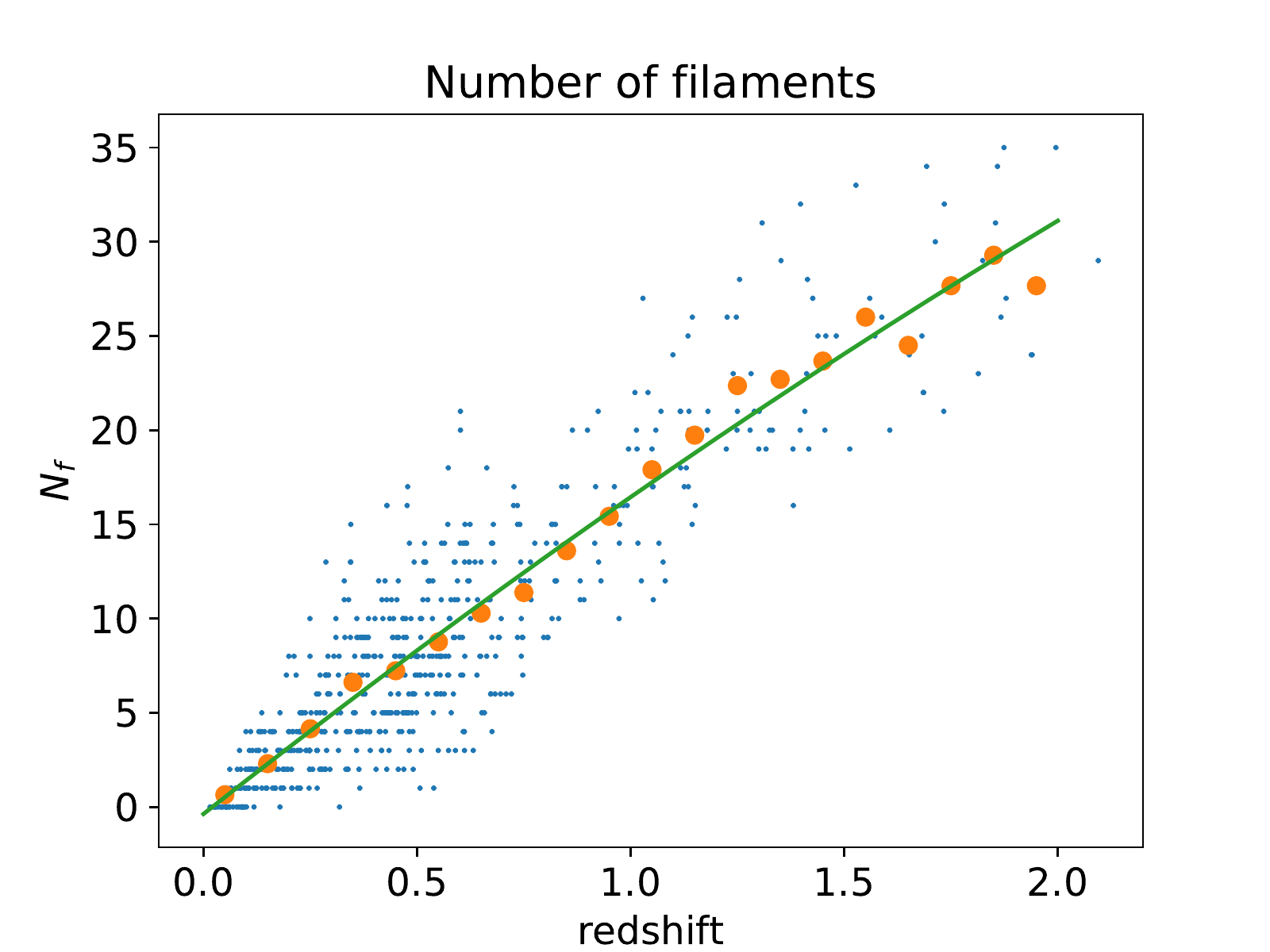}
   \caption{Number of filaments $N_f$ versus redshift of each source of the LoTSS sample that falls into the field covered by the filaments catalogues (dots). The best quadratic fit (solid line) and the mean in redshift bins (circles) are also reported.}
              \label{Fig:nfil}%
    \end{figure}

The depolarization is expected to follow a similar behaviour as described by Equation (\ref{eq:sigrm}) for the propagation of the polarized emission through cosmic web filaments, which can be written as: 
\begin{equation}
    \frac{p}{p_0}(z)=e^{-2\,\sigma_{\rm RRM_{0,f}}^2\,N_f(z)\,\lambda^4}
    \label{eq:sigrmf}
\end{equation}
where $\sigma_{\rm RRM_0,f}$ is the average $\sigma_{\rm RM}$ of a single filament and $N_f$ the number of filaments the radiation goes through. The term $p_0$ is taken from the linear fit of $p$ at $z=0$ (Section \ref{sec:rmvz_LoTSS}). We estimate the number of  filaments intercepted  by   each source of the LoTSS sample using the filaments catalogues of \citet{2015MNRAS.454.1140C} and \citet{2021arXiv210605253C} that extend out to $z=0.7$ and $z=2.2$, respectively. The catalogues cover the area of the Sloan Digital Sky Survey (SDSS) and 745 sources  of our LoTSS sample fall in it. We assume a typical filament transverse width of 3-Mpc \citep{2014MNRAS.441.2923C}. The number of filaments intercepted by the individual sources versus their redshift and their quadratic fit 
\begin{equation}
    N_f(z) = -1.08\,z^2   +    17.89\,z  -  0.37
    \label{eq:nffit}
\end{equation}
are shown in Figure \ref{Fig:nfil}. 

   \begin{figure*}
   \centering
    \includegraphics[width=\columnwidth]{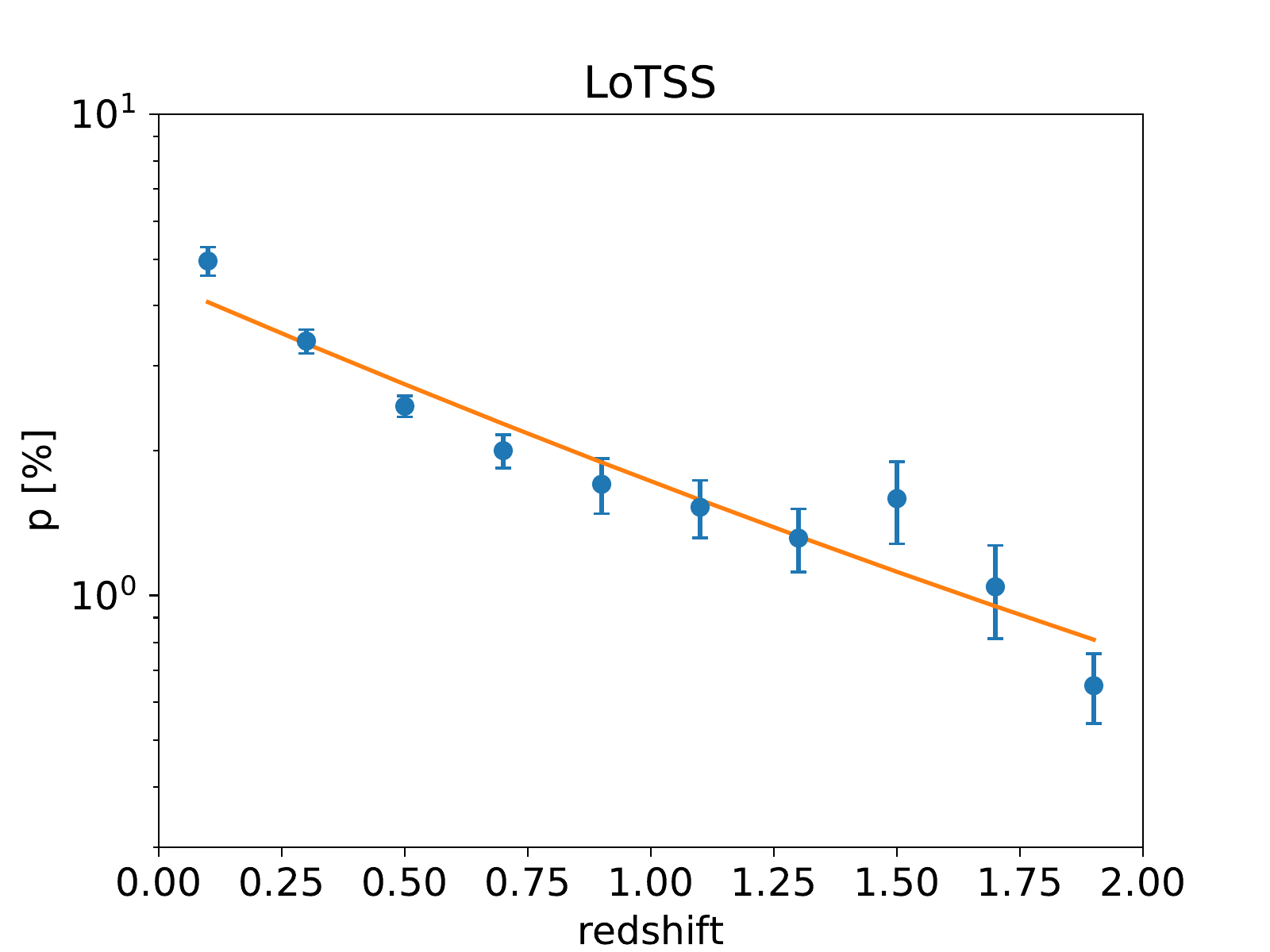}
    \includegraphics[width=\columnwidth]{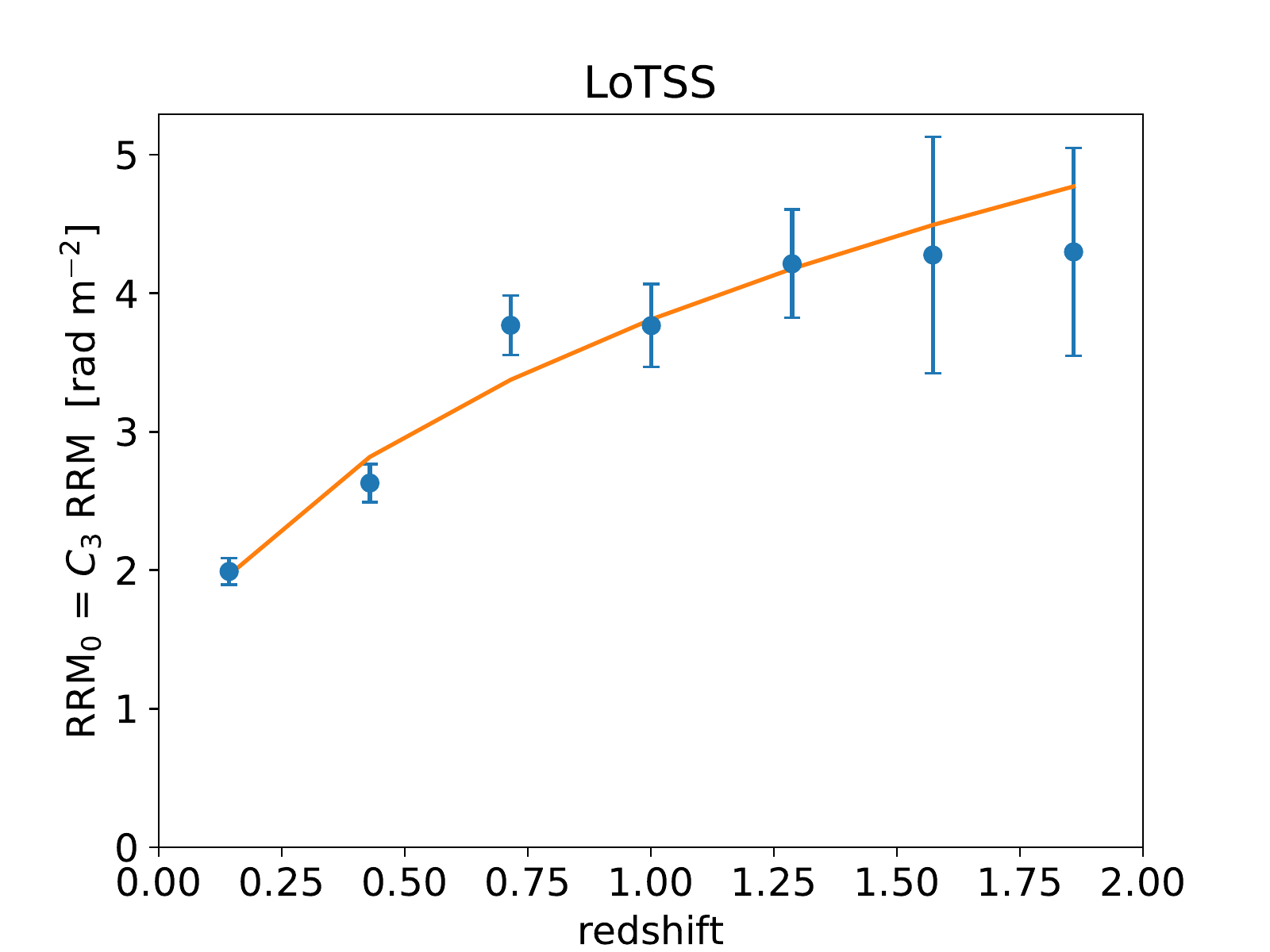}
   \caption{{\em Left}: Polarization fraction $p$ versus redshift of the LoTSS sample (circles) and  best fit of the function of Equation (\ref{eq:sigrmf}) (solid line). {\em Right}: RRM$_0$ as a function of redshift of the LoTSS sample corrected for model $C_3$ (circles) and best fit of the function of Equation (\ref{eq:rrmf}) (solid line).}
              \label{Fig:prmfit}%
    \end{figure*}

The best fit of Equation (\ref{eq:sigrmf}) to the depolarization  as a function of $z$ of the LoTSS sample is shown in Figure \ref{Fig:prmfit}, left panel. The curve fits the data well, further supporting the cosmic web filament origin of the depolarization. The best estimate of $\sigma_{\rm RRM_{0,f}}$ is
\begin{equation}
    \sigma_{\rm RRM_0,f} =0.0389 \pm 0.0010 \,\, \rm rad\,m^{-2}
\end{equation}
which gives an estimate of the average RM turbulence of filaments (the first to our knowledge). We regard this as an upper limit because part of the depolarization might be of a different origin. It is a conservative upper limit, for a cosmic filament origin is favoured and $ \sigma_{\rm RRM_0,f}$ is at least 50\% of our estimate.

If the RRM is generated by the filaments, then its rms deviation  is expected to be mostly flat with $z$ \citep[][]{2011ApJ...738..134A}, as we observe at 144-MHz, and  $\rm \left< RRM_0^2 \right>^{1/2}$  is expected to increase with redshift as $N_f^{1/2}(z)$ \citep{2010ApJ...723..476A}, because the path length through the filaments is a random walk and the RRM of each filament can be either negative or positive. The best redshift correction for filaments is $C_3$, as discussed in Section \ref{sec:rmvz_LoTSS}. We fit the RRM$_0$ rms values that we measure with equal-width redshift bins for the LoTSS sample with the function 
\begin{equation}
    {\rm RRM_0}(z) = {\rm RRM}_{0,f}\, \, N_f^{1/2}(z) + A_{\rm RRM}
    \label{eq:rrmf}
\end{equation}
where $ {\rm RRM}_{0,f}$ is the average absolute value of the RRM$_0$ of an individual filament, $N_f(z)$ is taken from  Equation (\ref{eq:nffit}), and $A_{RRM}$ is a constant term that accounts for possible other components of different origin. The resulting fit is shown in Figure \ref{Fig:prmfit}, right panel. The fit is a good approximation of the data, making the filament scenario self-consistent again, and the resulting filament RRM$_{0,f}$ is
\begin{equation}
    {\rm RRM}_{0,f} = 0.71 \pm 0.07  \,\, \rm rad\,m^{-2}
\end{equation}
in broad agreement with the value of  1.5-rad m$^{-2}$  from simulations \citep{2010ApJ...723..476A} and consistent with previous upper limits \citep[e.g., 3.8-rad m$^{-2}$ by ][]{2021MNRAS.503.2913A}. The constant term of the fit is $A_{\rm RRM}  = 0.91\pm 0.18  \,\, \rm rad\,m^{-2}$. 

Assuming a typical value for the electron density in filaments\footnote{$n_{e,f}$ is estimated at $z=0.7$ that  in terms of comoving distance is mid-way out to $z=2$, the range spanned by our data set.} of $n_{e,f}=10^{-5}$~cm$^{-3}$ \citep{2010ApJ...723..476A,2011ApJ...738..134A,2015A&A...580A.119V}, a filament width of 3-Mpc \citep{2014MNRAS.441.2923C}, and correcting it for the mean inclination of a filament to the line of sight (see Appendix \ref{app:len}), we get a filament magnetic field intensity parallel to the line of sight of
\begin{equation}
    B_{\parallel ,f} = 18.6\pm 1.9\,\, {\rm nG}
\end{equation}
and a filament total magnetic field of 
\begin{equation}
    B_f = \sqrt{3}\, B_{\parallel ,f}   =   32.3 \pm 3.2\,\, {\rm nG},
\end{equation}
assuming no dependence on $z$ of $n_e$ and the magnetic field $B$.  
This is in agreement with \citet{2021MNRAS.505.4178V} who found a magnetic field intensity per  filament of $30 \leq B_f \leq 60$-nG  using synchrotron emission stacking. It is also consistent with previous cosmic web  magnetic field upper limits or estimates from simulations. \citet{2017MNRAS.467.4914V} and \citet{2017MNRAS.468.4246B} found upper limits of 30--200-nG  from cross-correlating  synchrotron emission  with the large scale structure distribution, \citet{2018MNRAS.479..776V} estimated fields of 10--50-nG from simulations constrained by observations, \citet{2019ApJ...878...92V} estimated an upper limit of 40-nG with RMs of extragalactic source pairs, \citet{2021MNRAS.503.2913A} found an upper limit of 50-nG cross-correlating RMs with the galaxy distribution, \citet{2019A&A...622A..16O} found an upper limit of 250-nG from a differential number of filaments in the foreground of two lobes of a radio galaxy, and \citet{2021A&A...652A..80L} estimated an upper limit of 250-nG based on simulations constrained by non-detections.
 
 Our estimate is also in agreement with models based on primordial magnetic fields  amplified by astrophysical source seeding, which predict fields of a few tens of nG in filaments, in contrast to a few nG for models based on only  primordial magnetic fields grown  by MHD processes   \citep[e.g.,][]{2015A&A...580A.119V,2021MNRAS.505.5038A}. In principle, models based only on primordial magnetic fields can reach the measured amplitudes but only in the cases that the primordial seed field is at the top of the current upper limits \citep{2017CQGra..34w4001V}. In the more general case, a boost by astrophysical source seeding is required \citep[e.g.,][]{2015A&A...580A.119V,2017CQGra..34w4001V}.
 
 Our result of the detection of a filament RRM$_0$   supports  the presence of a baryonic Warm-Hot Ionised Medium (WHIM) in filaments, that  cosmological simulations predict to contain some 50\% of the cosmic baryons.
 
 We repeated the analysis fitting the RRMs measured in redshift bins with equal-number of sources obtaining consistent results (see Appendix \ref{app:B_iso}). 
 
 An estimate of the turbulent component of the magnetic field of a filament, assuming a Gaussian distribution, can be derived from  $\sigma_{\rm RRM_0,f}$ and the  Burn  Law  \citep{1966MNRAS.133...67B, 1996ASPC...88..271F, 1998MNRAS.299..189S, 2003A&A...401..835E, 1991MNRAS.253..147T,2004A&A...424..429M} with
\begin{eqnarray}
     \sigma_{\rm RRM_0,f} &=& 0.812 \,n_e \,\sigma_{B_{\parallel},f} \,\sqrt{l\,\lambda_B}\\
     B_{{\rm turb},f} &=& 2\,\sqrt{2/\pi} \,\sigma_{B_{\parallel},f} 
\end{eqnarray}
where $B_{{\rm turb},f}$ and $\sigma_{B_{\parallel},f} $ are the  mean turbulent magnetic field intensity and  the dispersion of  its  component parallel to the line of sight, $l$ is the radiation's path-length through the filament, and $\lambda_B$ is the coherence scale of the magnetic field.  The latter is expected to be a few $100 h^{-1}$ kpc \citep{2010ApJ...723..476A} and we   assume it  to range  within $0.4 < \lambda_B < 1.0$ Mpc. Assuming for the other terms the same values we used to  estimate  $B_f$, we find $3.5 \lesssim B_{{\rm turb},f} \lesssim 5.5$-nG. This is small compared to $B_f$ and so we conclude that the turbulent component is subdominant.

 \subsection{Local environment}
The other  option is an origin of RRM and $p$ local to the source for the 144-MHz data. 
The following considerations can be drawn from the analysis conducted in Section \ref{sec:disc_gc}: 
\begin{enumerate}
    \item The behaviour with redshift of $p$ at 144-MHz is still compatible with a local origin. The flat behaviour of RRM with $p$ instead disfavours it, because higher RRM values should be accompanied by stronger depolarization, as the 1.4-GHz data show.
    \item If the origin  occurs local to the source it is usually in the medium surrounding it \citep[e.g.][]{2008MNRAS.391..521L} and  increasing depolarization (i.e. decreasing $p$) might be related to larger turbulence at high redshift (Equation (\ref{eq:sigrm})).  The observed sources are in low density environments far from galaxy clusters, as we have shown, and thus have to be field or group galaxies that can be in a less relaxed state at high redshift than at present with higher turbulence. This can explain the behaviour at 1.4-GHz (see also \citealt{2021A&A...653A.155B}), but it would cause total depolarization at 144-MHz, which is not what we observe. 
\end{enumerate} 
We conclude that our data disfavour a local origin being the dominant factor for both the RRM and $p$ observed at 144-MHz, while it most likely is the dominant factor for the 1.4-GHz data (see Section \ref{sec:disc_gc}).  

%

\section{Conclusions}
\label{sec:conc}
We have  analysed the LoTSS DR2 RM Catalogue containing RMs measured at 144-MHz, and in particular a subset of sources with a spectroscopic redshift and above a Galactic latitude cut of $|b| = 25^\circ$, to study the behaviour with redshift of polarization quantities of extragalactic sources in low density environments in the range  $0 < z < 2$.  After subtracting the Galactic RM contribution and producing a catalogue of Residual RM (RRM), we measured the behaviour with redshift of the RRM rms deviation and  fractional polarization $p$. We also measured  $\left< {\rm RRM^2} \right>^{1/2}$ as a function of $p$. We  repeated the same analysis for NVSS RMs  of sources in the LoTSS sample, to investigate the behaviour of RRM and $p$ measured at 1.4-GHz of the same sample from low density environments. 

Our main findings  are: 
\begin{enumerate}
    \item At 144-MHz the RRM rms  is flat with redshift out to $z=2$. Once the redshift correction is applied, the rest frequency RRM$_0$ rms increases with redshift at a high confidence level for all of the  correction models we considered, showing a clear evolution with redshift. At 1.4-GHz  the RRM shows a hint of an increase with redshift and   RRM$_0$ increases with $z$ at a high confidence level.
    \item At 144-MHz the fractional polarization is anti-correlated with redshift,  at $z=1.9$ it is $\sim$1/8th of that at $z=0.1$, showing  a  evolution with redshift at high confidence level. Also at 1.4-GHz $p$ is   anti-correlated with redshift at a high confidence level, even though at a lower rate than at 144-MHz (at the high $z$ end it is $\sim$1/3rd of that at the low $z$ end).
    \item The RRM rms is flat with  $p$   at 144-MHz and no increase of RRM at low $p$ is observed. At 1.4-GHz, instead, RRM rms  decreases with $p$, indicating that sources with a higher RRM are more depolarized.   
\end{enumerate}

These findings and our analysis lead us to  the following  main results:  
\begin{enumerate}
    \item There is a clear evolution with redshift of $p$ and  the rest frame RRM$_0$ for the physically-motivated redshift correction models we considered.
    \item Polarized sources at 144-MHz reside far from galaxy clusters with  a peak at $\sim 5\,R_{200}$, confirming they are in low density environments.  The general 1.4-GHz population is closer to clusters, instead, with a substantial fraction within  cluster boundaries. 
    \item The RRM and $p(z)$ have a different origin at 144-MHz and 1.4-GHz. Depolarization at 1.4-GHz is not mainly due to the radiation travelling through the IGM on large scales and  a local origin is favoured for the RRM and depolarization at this frequency.  A passage through filaments of the cosmic web is favoured as the origin of the RRM and $p(z)$ at 144-MHz. The depolarization with $z$, the flat behaviour of the RRM with $z$ and $p$,  and the fit of $p$ and RRM$_0$ to that expected by the number of filaments along the line of sight are all consistent with such an origin. 
    \item If we attribute the total RRM and $p(z)$ to cosmic web filaments, we estimate an average RM for an individual filament of  ${\rm RRM}_{0,f} = 0.71 \pm 0.07  \,\, \rm rad\,m^{-2}$, and an average magnetic field per filament of $  B_f = 32 \pm 3\,\, {\rm nG}$, assuming no dependence on $z$ of $n_e$ and $B$. This value favours models where the field in filaments is amplified by astrophysical source seeding in contrast with models solely based on the growth of  primordial magnetic fields.
    \item The detection of a filament RRM$_0$ supports  the presence of a diffuse WHIM in cosmic filaments that were predicted to contain $\sim 50$\% of the cosmic baryons.
    \item We also estimate, for the first time,  an average turbulence in each filament of $\sigma_{\rm RRM_{0,f}} =0.0389 \pm 0.0010 \,\, \rm rad\,m^{-2}$. We use it to estimate the turbulent component of the magnetic field in filaments and find that it is subdominant.
\end{enumerate}
With this work we have applied RMs measured at low radio frequencies to detect and measure magnetic fields in  cosmic web filaments.  We have several hints that the bulk of what we observe at 144-MHz is  generated in the IGM, while at 1.4-GHz the observations are dominated by the local source environment. Our estimate for the magnetic field in filaments is in agreement with that found using an independent method based on synchrotron emission stacking \citep{2021MNRAS.505.4178V}, indicating that these are complementary and effective methods to investigate the cosmic web magnetism.  This work shows the importance of low frequency observations and the availability of sources with measured redshifts to investigate the magnetism in the cosmic web and its evolution with cosmic time. It  also shows the importance of having RM measurements at different frequencies to discriminate their origin. 

This is an important  first step. Several new advances  can be achieved in future work. Modelling of the behaviour with $z$ of $n_e$ and $B_f$ can improve our estimate of the filament magnetic field and turbulence, and possibly estimate their evolution with redshift. Larger samples will improve the redshift resolution and bin sensitivity, which will give a more detailed view of the evolution with redshift and  extend it to higher redshifts. It will also allow separation into different source populations and investigations of their impact and their own evolution.  A first step will be achieved by completing the LoTSS survey and a further leap can be made with a polarization survey carried out with SKA1-LOW \citep{2019arXiv191212699B}. Comparing results at different frequencies has been essential to the determination of where the RRM is originating. Combining  LoTSS, ASKAP-POSSUM \citep{2010AAS...21547013G} and APERTIF \citep{2021A&A...653A.155B} data, and in the future  those of  the surveys of SKA1--LOW and SKA1--MID \citep{2019arXiv191212699B}, will help establish more firmly  the RRM origin  at several frequencies and at what frequency the IGM component starts to prevail. 
 The same surveys can also be used to improve the separation of the extragalactic and Galactic RM components \citep{2022A&A...657A..43H}, which is needed to improve the overall precision of the RRM estimates.

\section*{Acknowledgements}

This work has been conducted   within the LOFAR Magnetism Key Science Project\footnote{https://lofar-mksp.org/} (MKSP). 
This work has made use of LoTSS DR2 data. 
VV acknowledges support from INAF mainstream project “Galaxy Clusters Science with LOFAR” 1.05.01.86.05. 
AMS gratefully acknowledges support from the Alan Turing Institute, grant reference EP/V030302/1.
CS acknowledges support from the MIUR grant FARE ``SMS".
LOFAR \citep{2013A&A...556A...2V} is the Low Frequency Array designed and constructed by ASTRON. It has observing, data processing, and data storage facilities in several countries, which are owned by various parties (each with their own funding sources), and which are collectively operated by the ILT foundation under a joint scientific policy. The ILT resources have benefited from the following recent major funding sources: CNRS-INSU, Observatoire de Paris and Universit\'e d'Orl\'eans, France; BMBF, MIWF-NRW, MPG, Germany; Science Foundation Ireland (SFI), Department of Business, Enterprise and Innovation (DBEI), Ireland; NWO, The Netherlands; The Science and Technology Facilities Council, UK; Ministry of Science and Higher Education, Poland; The Istituto Nazionale di Astrofisica (INAF), Italy.   LoTSS made use of the Dutch national e-infrastructure with support of the SURF Cooperative (e-infra 180169) and the LOFAR e-infra group. The J\"ulich LOFAR Long Term Archive and the German LOFAR network are both coordinated and operated by the J\"ulich Supercomputing Centre (JSC), and computing resources on the supercomputer JUWELS at JSC were provided by the Gauss Centre for Supercomputing e.V. (grant CHTB00) through the John von Neumann Institute for Computing (NIC).
LoTSS made use of the University of Hertfordshire high-performance computing facility and the LOFAR-UK computing facility located at the University of Hertfordshire and supported by STFC [ST/P000096/1], and of the Italian LOFAR IT computing infrastructure supported and operated by INAF, and by the Physics Department of Turin university (under an agreement with Consorzio Interuniversitario per la Fisica Spaziale) at the C3S Supercomputing Centre, Italy.
This work made use  of the Python packages Astropy \citep{2013A&A...558A..33A}  and Matplotlib \citep{2007CSE.....9...90H}. Some of the results in this paper have been derived using the healpy \citep{2019JOSS....4.1298Z} and HEALPix\footnote{http://healpix.sf.net} \citep{2005ApJ...622..759G} packages. 
\section*{Data Availability}

The LoTSS DR2 RM catalogue  will be publicly released as the  paper describing it will be accepted for publication. 
 The Hammond catalogue of NVSS RMs with  redshift cross-matches \citep{2012arXiv1209.1438H}, the NVSS RM catalogue \citep{2009ApJ...702.1230T}, the galaxy cluster catalogue \citep{2015ApJ...807..178W}, and the cosmic web filament catalogues \citep{2016MNRAS.461.3896C,2021arXiv210605253C} are all available at the web sites reported in the papers that describe them.



\bibliographystyle{mnras}
\bibliography{rmvz} 

\begin{thebibliography}{}
\makeatletter
\relax
\def\mn@urlcharsother{\let\do\@makeother \do\$\do\&\do\#\do\^\do\_\do\%\do\~}
\def\mn@doi{\begingroup\mn@urlcharsother \@ifnextchar [ {\mn@doi@}
  {\mn@doi@[]}}
\def\mn@doi@[#1]#2{\def\@tempa{#1}\ifx\@tempa\@empty \href
  {http://dx.doi.org/#2} {doi:#2}\else \href {http://dx.doi.org/#2} {#1}\fi
  \endgroup}
\def\mn@eprint#1#2{\mn@eprint@#1:#2::\@nil}
\def\mn@eprint@arXiv#1{\href {http://arxiv.org/abs/#1} {{\tt arXiv:#1}}}
\def\mn@eprint@dblp#1{\href {http://dblp.uni-trier.de/rec/bibtex/#1.xml}
  {dblp:#1}}
\def\mn@eprint@#1:#2:#3:#4\@nil{\def\@tempa {#1}\def\@tempb {#2}\def\@tempc
  {#3}\ifx \@tempc \@empty \let \@tempc \@tempb \let \@tempb \@tempa \fi \ifx
  \@tempb \@empty \def\@tempb {arXiv}\fi \@ifundefined
  {mn@eprint@\@tempb}{\@tempb:\@tempc}{\expandafter \expandafter \csname
  mn@eprint@\@tempb\endcsname \expandafter{\@tempc}}}

\bibitem[\protect\citeauthoryear{{Akahori} \& {Ryu}}{{Akahori} \&
  {Ryu}}{2010}]{2010ApJ...723..476A}
{Akahori} T.,  {Ryu} D.,  2010, \mn@doi [\apj] {10.1088/0004-637X/723/1/476},
  \href {https://ui.adsabs.harvard.edu/abs/2010ApJ...723..476A} {723, 476}

\bibitem[\protect\citeauthoryear{{Akahori} \& {Ryu}}{{Akahori} \&
  {Ryu}}{2011}]{2011ApJ...738..134A}
{Akahori} T.,  {Ryu} D.,  2011, \mn@doi [\apj] {10.1088/0004-637X/738/2/134},
  \href {https://ui.adsabs.harvard.edu/abs/2011ApJ...738..134A} {738, 134}

\bibitem[\protect\citeauthoryear{{Amaral}, {Vernstrom}  \& {Gaensler}}{{Amaral}
  et~al.}{2021}]{2021MNRAS.503.2913A}
{Amaral} A.~D.,  {Vernstrom} T.,   {Gaensler} B.~M.,  2021, \mn@doi [\mnras]
  {10.1093/mnras/stab564}, \href
  {https://ui.adsabs.harvard.edu/abs/2021MNRAS.503.2913A} {503, 2913}

\bibitem[\protect\citeauthoryear{{Ar{\'a}mburo-Garc{\'\i}a}, {Bondarenko},
  {Boyarsky}, {Nelson}, {Pillepich}  \& {Sokolenko}}{{Ar{\'a}mburo-Garc{\'\i}a}
  et~al.}{2021}]{2021MNRAS.505.5038A}
{Ar{\'a}mburo-Garc{\'\i}a} A.,  {Bondarenko} K.,  {Boyarsky} A.,  {Nelson} D.,
  {Pillepich} A.,   {Sokolenko} A.,  2021, \mn@doi [\mnras]
  {10.1093/mnras/stab1632}, \href
  {https://ui.adsabs.harvard.edu/abs/2021MNRAS.505.5038A} {505, 5038}

\bibitem[\protect\citeauthoryear{{Astropy Collaboration} et~al.,}{{Astropy
  Collaboration} et~al.}{2013}]{2013A&A...558A..33A}
{Astropy Collaboration} et~al., 2013, \mn@doi [\aap]
  {10.1051/0004-6361/201322068}, \href
  {https://ui.adsabs.harvard.edu/abs/2013A&A...558A..33A} {558, A33}

\bibitem[\protect\citeauthoryear{{Berger}, {Adebahr}, {Herrera Ruiz}, {Wright},
  {Prandoni}  \& {Dettmar}}{{Berger} et~al.}{2021}]{2021A&A...653A.155B}
{Berger} A.,  {Adebahr} B.,  {Herrera Ruiz} N.,  {Wright} A.~H.,  {Prandoni}
  I.,   {Dettmar} R.~J.,  2021, \mn@doi [\aap] {10.1051/0004-6361/202040009},
  \href {https://ui.adsabs.harvard.edu/abs/2021A&A...653A.155B} {653, A155}

\bibitem[\protect\citeauthoryear{{Bernet}, {Miniati}, {Lilly}, {Kronberg}  \&
  {Dessauges-Zavadsky}}{{Bernet} et~al.}{2008}]{2008Natur.454..302B}
{Bernet} M.~L.,  {Miniati} F.,  {Lilly} S.~J.,  {Kronberg} P.~P.,
  {Dessauges-Zavadsky} M.,  2008, \mn@doi [\nat] {10.1038/nature07105}, \href
  {https://ui.adsabs.harvard.edu/abs/2008Natur.454..302B} {454, 302}

\bibitem[\protect\citeauthoryear{{Braun}, {Bonaldi}, {Bourke}, {Keane}  \&
  {Wagg}}{{Braun} et~al.}{2019}]{2019arXiv191212699B}
{Braun} R.,  {Bonaldi} A.,  {Bourke} T.,  {Keane} E.,   {Wagg} J.,  2019, arXiv
  e-prints, \href {https://ui.adsabs.harvard.edu/abs/2019arXiv191212699B} {p.
  arXiv:1912.12699}

\bibitem[\protect\citeauthoryear{{Brentjens} \& {de Bruyn}}{{Brentjens} \& {de
  Bruyn}}{2005}]{2005A&A...441.1217B}
{Brentjens} M.~A.,  {de Bruyn} A.~G.,  2005, \mn@doi [\aap]
  {10.1051/0004-6361:20052990}, \href
  {https://ui.adsabs.harvard.edu/abs/2005A&A...441.1217B} {441, 1217}

\bibitem[\protect\citeauthoryear{{Brown} et~al.,}{{Brown}
  et~al.}{2017}]{2017MNRAS.468.4246B}
{Brown} S.,  et~al., 2017, \mn@doi [\mnras] {10.1093/mnras/stx746}, \href
  {https://ui.adsabs.harvard.edu/abs/2017MNRAS.468.4246B} {468, 4246}

\bibitem[\protect\citeauthoryear{{Burn}}{{Burn}}{1966}]{1966MNRAS.133...67B}
{Burn} B.~J.,  1966, \mn@doi [\mnras] {10.1093/mnras/133.1.67}, \href
  {https://ui.adsabs.harvard.edu/abs/1966MNRAS.133...67B} {133, 67}

\bibitem[\protect\citeauthoryear{{Carr{\'o}n Duque}, {Migliaccio}, {Marinucci}
  \& {Vittorio}}{{Carr{\'o}n Duque} et~al.}{2021}]{2021arXiv210605253C}
{Carr{\'o}n Duque} J.,  {Migliaccio} M.,  {Marinucci} D.,   {Vittorio} N.,
  2021, arXiv e-prints, \href
  {https://ui.adsabs.harvard.edu/abs/2021arXiv210605253C} {p. arXiv:2106.05253}

\bibitem[\protect\citeauthoryear{{Cautun}, {van de Weygaert}, {Jones}  \&
  {Frenk}}{{Cautun} et~al.}{2014}]{2014MNRAS.441.2923C}
{Cautun} M.,  {van de Weygaert} R.,  {Jones} B. J.~T.,   {Frenk} C.~S.,  2014,
  \mn@doi [\mnras] {10.1093/mnras/stu768}, \href
  {https://ui.adsabs.harvard.edu/abs/2014MNRAS.441.2923C} {441, 2923}

\bibitem[\protect\citeauthoryear{{Chen}, {Ho}, {Freeman}, {Genovese}  \&
  {Wasserman}}{{Chen} et~al.}{2015}]{2015MNRAS.454.1140C}
{Chen} Y.-C.,  {Ho} S.,  {Freeman} P.~E.,  {Genovese} C.~R.,   {Wasserman} L.,
  2015, \mn@doi [\mnras] {10.1093/mnras/stv1996}, \href
  {https://ui.adsabs.harvard.edu/abs/2015MNRAS.454.1140C} {454, 1140}

\bibitem[\protect\citeauthoryear{{Chen}, {Ho}, {Brinkmann}, {Freeman},
  {Genovese}, {Schneider}  \& {Wasserman}}{{Chen}
  et~al.}{2016}]{2016MNRAS.461.3896C}
{Chen} Y.-C.,  {Ho} S.,  {Brinkmann} J.,  {Freeman} P.~E.,  {Genovese} C.~R.,
  {Schneider} D.~P.,   {Wasserman} L.,  2016, \mn@doi [\mnras]
  {10.1093/mnras/stw1554}, \href
  {https://ui.adsabs.harvard.edu/abs/2016MNRAS.461.3896C} {461, 3896}

\bibitem[\protect\citeauthoryear{{En{\ss}lin} \& {Vogt}}{{En{\ss}lin} \&
  {Vogt}}{2003}]{2003A&A...401..835E}
{En{\ss}lin} T.~A.,  {Vogt} C.,  2003, \mn@doi [\aap]
  {10.1051/0004-6361:20030172}, \href
  {https://ui.adsabs.harvard.edu/abs/2003A&A...401..835E} {401, 835}

\bibitem[\protect\citeauthoryear{{Ettori} \& {Balestra}}{{Ettori} \&
  {Balestra}}{2009}]{2009A&A...496..343E}
{Ettori} S.,  {Balestra} I.,  2009, \mn@doi [\aap]
  {10.1051/0004-6361:200811177}, \href
  {https://ui.adsabs.harvard.edu/abs/2009A&A...496..343E} {496, 343}

\bibitem[\protect\citeauthoryear{{Felten}}{{Felten}}{1996}]{1996ASPC...88..271F}
{Felten} J.~E.,  1996, in {Trimble} V.,  {Reisenegger} A.,  eds,  Astronomical
  Society of the Pacific Conference Series Vol. 88, Clusters, Lensing, and the
  Future of the Universe. p.~271

\bibitem[\protect\citeauthoryear{{Fujimoto}, {Kawabata}  \& {Sofue}}{{Fujimoto}
  et~al.}{1971}]{1971PThPS..49..181F}
{Fujimoto} M.,  {Kawabata} K.,   {Sofue} Y.,  1971, \mn@doi [Progress of
  Theoretical Physics Supplement] {10.1143/PTPS.49.181}, \href
  {https://ui.adsabs.harvard.edu/abs/1971PThPS..49..181F} {49, 181}

\bibitem[\protect\citeauthoryear{{Gaensler}, {Landecker}, {Taylor}  \& {POSSUM
  Collaboration}}{{Gaensler} et~al.}{2010}]{2010AAS...21547013G}
{Gaensler} B.~M.,  {Landecker} T.~L.,  {Taylor} A.~R.,   {POSSUM Collaboration}
  2010, in American Astronomical Society Meeting Abstracts \#215. p. 470.13

\bibitem[\protect\citeauthoryear{{Gheller}, {Vazza}, {Favre}  \&
  {Br{\"u}ggen}}{{Gheller} et~al.}{2015}]{2015MNRAS.453.1164G}
{Gheller} C.,  {Vazza} F.,  {Favre} J.,   {Br{\"u}ggen} M.,  2015, \mn@doi
  [\mnras] {10.1093/mnras/stv1646}, \href
  {https://ui.adsabs.harvard.edu/abs/2015MNRAS.453.1164G} {453, 1164}

\bibitem[\protect\citeauthoryear{{G{\'o}rski}, {Hivon}, {Banday}, {Wandelt},
  {Hansen}, {Reinecke}  \& {Bartelmann}}{{G{\'o}rski}
  et~al.}{2005}]{2005ApJ...622..759G}
{G{\'o}rski} K.~M.,  {Hivon} E.,  {Banday} A.~J.,  {Wandelt} B.~D.,  {Hansen}
  F.~K.,  {Reinecke} M.,   {Bartelmann} M.,  2005, \mn@doi [\apj]
  {10.1086/427976}, \href
  {https://ui.adsabs.harvard.edu/abs/2005ApJ...622..759G} {622, 759}

\bibitem[\protect\citeauthoryear{{Govoni} et~al.,}{{Govoni}
  et~al.}{2019}]{2019Sci...364..981G}
{Govoni} F.,  et~al., 2019, \mn@doi [Science] {10.1126/science.aat7500}, \href
  {https://ui.adsabs.harvard.edu/abs/2019Sci...364..981G} {364, 981}

\bibitem[\protect\citeauthoryear{{Hammond}, {Robishaw}  \&
  {Gaensler}}{{Hammond} et~al.}{2012}]{2012arXiv1209.1438H}
{Hammond} A.~M.,  {Robishaw} T.,   {Gaensler} B.~M.,  2012, arXiv e-prints,
  \href {https://ui.adsabs.harvard.edu/abs/2012arXiv1209.1438H} {p.
  arXiv:1209.1438}

\bibitem[\protect\citeauthoryear{{Heald} et~al.,}{{Heald}
  et~al.}{2020}]{2020Galax...8...53H}
{Heald} G.,  et~al., 2020, \mn@doi [Galaxies] {10.3390/galaxies8030053}, \href
  {https://ui.adsabs.harvard.edu/abs/2020Galax...8...53H} {8, 53}

\bibitem[\protect\citeauthoryear{{Hunter}}{{Hunter}}{2007}]{2007CSE.....9...90H}
{Hunter} J.~D.,  2007, \mn@doi [Computing in Science and Engineering]
  {10.1109/MCSE.2007.55}, \href
  {https://ui.adsabs.harvard.edu/abs/2007CSE.....9...90H} {9, 90}

\bibitem[\protect\citeauthoryear{{Hutschenreuter} et~al.,}{{Hutschenreuter}
  et~al.}{2022}]{2022A&A...657A..43H}
{Hutschenreuter} S.,  et~al., 2022, \mn@doi [\aap]
  {10.1051/0004-6361/202140486}, \href
  {https://ui.adsabs.harvard.edu/abs/2022A&A...657A..43H} {657, A43}

\bibitem[\protect\citeauthoryear{{Kronberg} \& {Perry}}{{Kronberg} \&
  {Perry}}{1982}]{1982ApJ...263..518K}
{Kronberg} P.~P.,  {Perry} J.~J.,  1982, \mn@doi [\apj] {10.1086/160523}, \href
  {https://ui.adsabs.harvard.edu/abs/1982ApJ...263..518K} {263, 518}

\bibitem[\protect\citeauthoryear{{Kronberg}, {Reinhardt}  \&
  {Simard-Normandin}}{{Kronberg} et~al.}{1977}]{1977A&A....61..771K}
{Kronberg} P.~P.,  {Reinhardt} M.,   {Simard-Normandin} M.,  1977, \aap, \href
  {https://ui.adsabs.harvard.edu/abs/1977A&A....61..771K} {61, 771}

\bibitem[\protect\citeauthoryear{{Kronberg}, {Bernet}, {Miniati}, {Lilly},
  {Short}  \& {Higdon}}{{Kronberg} et~al.}{2008}]{2008ApJ...676...70K}
{Kronberg} P.~P.,  {Bernet} M.~L.,  {Miniati} F.,  {Lilly} S.~J.,  {Short}
  M.~B.,   {Higdon} D.~M.,  2008, \mn@doi [\apj] {10.1086/527281}, \href
  {https://ui.adsabs.harvard.edu/abs/2008ApJ...676...70K} {676, 70}

\bibitem[\protect\citeauthoryear{{Laing}, {Bridle}, {Parma}  \&
  {Murgia}}{{Laing} et~al.}{2008}]{2008MNRAS.391..521L}
{Laing} R.~A.,  {Bridle} A.~H.,  {Parma} P.,   {Murgia} M.,  2008, \mn@doi
  [\mnras] {10.1111/j.1365-2966.2008.13895.x}, \href
  {https://ui.adsabs.harvard.edu/abs/2008MNRAS.391..521L} {391, 521}

\bibitem[\protect\citeauthoryear{{Lamee}, {Rudnick}, {Farnes}, {Carretti},
  {Gaensler}, {Haverkorn}  \& {Poppi}}{{Lamee}
  et~al.}{2016}]{2016ApJ...829....5L}
{Lamee} M.,  {Rudnick} L.,  {Farnes} J.~S.,  {Carretti} E.,  {Gaensler} B.~M.,
  {Haverkorn} M.,   {Poppi} S.,  2016, \mn@doi [\apj]
  {10.3847/0004-637X/829/1/5}, \href
  {https://ui.adsabs.harvard.edu/abs/2016ApJ...829....5L} {829, 5}

\bibitem[\protect\citeauthoryear{{Locatelli}, {Vazza}, {Bonafede}, {Banfi},
  {Bernardi}, {Gheller}, {Botteon}  \& {Shimwell}}{{Locatelli}
  et~al.}{2021}]{2021A&A...652A..80L}
{Locatelli} N.,  {Vazza} F.,  {Bonafede} A.,  {Banfi} S.,  {Bernardi} G.,
  {Gheller} C.,  {Botteon} A.,   {Shimwell} T.,  2021, \mn@doi [\aap]
  {10.1051/0004-6361/202140526}, \href
  {https://ui.adsabs.harvard.edu/abs/2021A&A...652A..80L} {652, A80}

\bibitem[\protect\citeauthoryear{{Murgia}, {Govoni}, {Feretti}, {Giovannini},
  {Dallacasa}, {Fanti}, {Taylor}  \& {Dolag}}{{Murgia}
  et~al.}{2004}]{2004A&A...424..429M}
{Murgia} M.,  {Govoni} F.,  {Feretti} L.,  {Giovannini} G.,  {Dallacasa} D.,
  {Fanti} R.,  {Taylor} G.~B.,   {Dolag} K.,  2004, \mn@doi [\aap]
  {10.1051/0004-6361:20040191}, \href
  {https://ui.adsabs.harvard.edu/abs/2004A&A...424..429M} {424, 429}

\bibitem[\protect\citeauthoryear{{O'Sullivan} et~al.,}{{O'Sullivan}
  et~al.}{2019}]{2019A&A...622A..16O}
{O'Sullivan} S.~P.,  et~al., 2019, \mn@doi [\aap]
  {10.1051/0004-6361/201833832}, \href
  {https://ui.adsabs.harvard.edu/abs/2019A&A...622A..16O} {622, A16}

\bibitem[\protect\citeauthoryear{{O'Sullivan} et~al.,}{{O'Sullivan}
  et~al.}{2020}]{2020MNRAS.495.2607O}
{O'Sullivan} S.~P.,  et~al., 2020, \mn@doi [\mnras] {10.1093/mnras/staa1395},
  \href {https://ui.adsabs.harvard.edu/abs/2020MNRAS.495.2607O} {495, 2607}

\bibitem[\protect\citeauthoryear{{Oppermann} et~al.,}{{Oppermann}
  et~al.}{2015}]{2015A&A...575A.118O}
{Oppermann} N.,  et~al., 2015, \mn@doi [\aap] {10.1051/0004-6361/201423995},
  \href {https://ui.adsabs.harvard.edu/abs/2015A&A...575A.118O} {575, A118}

\bibitem[\protect\citeauthoryear{{Oren} \& {Wolfe}}{{Oren} \&
  {Wolfe}}{1995}]{1995ApJ...445..624O}
{Oren} A.~L.,  {Wolfe} A.~M.,  1995, \mn@doi [\apj] {10.1086/175726}, \href
  {https://ui.adsabs.harvard.edu/abs/1995ApJ...445..624O} {445, 624}

\bibitem[\protect\citeauthoryear{{Planck Collaboration} et~al.,}{{Planck
  Collaboration} et~al.}{2020}]{2020A&A...641A...6P}
{Planck Collaboration} et~al., 2020, \mn@doi [\aap]
  {10.1051/0004-6361/201833910}, \href
  {https://ui.adsabs.harvard.edu/abs/2020A&A...641A...6P} {641, A6}

\bibitem[\protect\citeauthoryear{{Porayko} et~al.,}{{Porayko}
  et~al.}{2019}]{2019MNRAS.483.4100P}
{Porayko} N.~K.,  et~al., 2019, \mn@doi [\mnras] {10.1093/mnras/sty3324}, \href
  {https://ui.adsabs.harvard.edu/abs/2019MNRAS.483.4100P} {483, 4100}

\bibitem[\protect\citeauthoryear{{Reinhardt}}{{Reinhardt}}{1972}]{1972A&A....19..104R}
{Reinhardt} M.,  1972, \aap, \href
  {https://ui.adsabs.harvard.edu/abs/1972A&A....19..104R} {19, 104}

\bibitem[\protect\citeauthoryear{{Riseley} et~al.,}{{Riseley}
  et~al.}{2020}]{2020PASA...37...29R}
{Riseley} C.~J.,  et~al., 2020, \mn@doi [\pasa] {10.1017/pasa.2020.20}, \href
  {https://ui.adsabs.harvard.edu/abs/2020PASA...37...29R} {37, e029}

\bibitem[\protect\citeauthoryear{{Shimwell} et~al.,}{{Shimwell}
  et~al.}{2017}]{2017A&A...598A.104S}
{Shimwell} T.~W.,  et~al., 2017, \mn@doi [\aap] {10.1051/0004-6361/201629313},
  \href {https://ui.adsabs.harvard.edu/abs/2017A&A...598A.104S} {598, A104}

\bibitem[\protect\citeauthoryear{{Shimwell} et~al.,}{{Shimwell}
  et~al.}{2019}]{2019A&A...622A...1S}
{Shimwell} T.~W.,  et~al., 2019, \mn@doi [\aap] {10.1051/0004-6361/201833559},
  \href {https://ui.adsabs.harvard.edu/abs/2019A&A...622A...1S} {622, A1}

\bibitem[\protect\citeauthoryear{{Sofue}, {Fujimoto}  \& {Kawabata}}{{Sofue}
  et~al.}{1979}]{1979PASJ...31..125S}
{Sofue} Y.,  {Fujimoto} M.,   {Kawabata} K.,  1979, \pasj, \href
  {https://ui.adsabs.harvard.edu/abs/1979PASJ...31..125S} {31, 125}

\bibitem[\protect\citeauthoryear{{Sokoloff}, {Bykov}, {Shukurov},
  {Berkhuijsen}, {Beck}  \& {Poezd}}{{Sokoloff}
  et~al.}{1998}]{1998MNRAS.299..189S}
{Sokoloff} D.~D.,  {Bykov} A.~A.,  {Shukurov} A.,  {Berkhuijsen} E.~M.,  {Beck}
  R.,   {Poezd} A.~D.,  1998, \mn@doi [\mnras]
  {10.1046/j.1365-8711.1998.01782.x}, \href
  {https://ui.adsabs.harvard.edu/abs/1998MNRAS.299..189S} {299, 189}

\bibitem[\protect\citeauthoryear{{Sotomayor-Beltran}
  et~al.,}{{Sotomayor-Beltran} et~al.}{2013}]{2013A&A...552A..58S}
{Sotomayor-Beltran} C.,  et~al., 2013, \mn@doi [\aap]
  {10.1051/0004-6361/201220728}, \href
  {https://ui.adsabs.harvard.edu/abs/2013A&A...552A..58S} {552, A58}

\bibitem[\protect\citeauthoryear{{Stuardi} et~al.,}{{Stuardi}
  et~al.}{2020}]{2020A&A...638A..48S}
{Stuardi} C.,  et~al., 2020, \mn@doi [\aap] {10.1051/0004-6361/202037635},
  \href {https://ui.adsabs.harvard.edu/abs/2020A&A...638A..48S} {638, A48}

\bibitem[\protect\citeauthoryear{{Subramanian}}{{Subramanian}}{2016}]{2016RPPh...79g6901S}
{Subramanian} K.,  2016, \mn@doi [Reports on Progress in Physics]
  {10.1088/0034-4885/79/7/076901}, \href
  {https://ui.adsabs.harvard.edu/abs/2016RPPh...79g6901S} {79, 076901}

\bibitem[\protect\citeauthoryear{{Taylor}, {Stil}  \& {Sunstrum}}{{Taylor}
  et~al.}{2009}]{2009ApJ...702.1230T}
{Taylor} A.~R.,  {Stil} J.~M.,   {Sunstrum} C.,  2009, \mn@doi [\apj]
  {10.1088/0004-637X/702/2/1230}, \href
  {https://ui.adsabs.harvard.edu/abs/2009ApJ...702.1230T} {702, 1230}

\bibitem[\protect\citeauthoryear{{Thomson} \& {Nelson}}{{Thomson} \&
  {Nelson}}{1982}]{1982MNRAS.201..365T}
{Thomson} R.~C.,  {Nelson} A.~H.,  1982, \mn@doi [\mnras]
  {10.1093/mnras/201.2.365}, \href
  {https://ui.adsabs.harvard.edu/abs/1982MNRAS.201..365T} {201, 365}

\bibitem[\protect\citeauthoryear{{Tribble}}{{Tribble}}{1991}]{1991MNRAS.253..147T}
{Tribble} P.~C.,  1991, \mn@doi [\mnras] {10.1093/mnras/253.1.147}, \href
  {https://ui.adsabs.harvard.edu/abs/1991MNRAS.253..147T} {253, 147}

\bibitem[\protect\citeauthoryear{{Vacca} et~al.,}{{Vacca}
  et~al.}{2018}]{2018MNRAS.479..776V}
{Vacca} V.,  et~al., 2018, \mn@doi [\mnras] {10.1093/mnras/sty1151}, \href
  {https://ui.adsabs.harvard.edu/abs/2018MNRAS.479..776V} {479, 776}

\bibitem[\protect\citeauthoryear{{Vazza}, {Ferrari}, {Br{\"u}ggen}, {Bonafede},
  {Gheller}  \& {Wang}}{{Vazza} et~al.}{2015}]{2015A&A...580A.119V}
{Vazza} F.,  {Ferrari} C.,  {Br{\"u}ggen} M.,  {Bonafede} A.,  {Gheller} C.,
  {Wang} P.,  2015, \mn@doi [\aap] {10.1051/0004-6361/201526228}, \href
  {https://ui.adsabs.harvard.edu/abs/2015A&A...580A.119V} {580, A119}

\bibitem[\protect\citeauthoryear{{Vazza}, {Br{\"u}ggen}, {Gheller},
  {Hackstein}, {Wittor}  \& {Hinz}}{{Vazza} et~al.}{2017}]{2017CQGra..34w4001V}
{Vazza} F.,  {Br{\"u}ggen} M.,  {Gheller} C.,  {Hackstein} S.,  {Wittor} D.,
  {Hinz} P.~M.,  2017, \mn@doi [Classical and Quantum Gravity]
  {10.1088/1361-6382/aa8e60}, \href
  {https://ui.adsabs.harvard.edu/abs/2017CQGra..34w4001V} {34, 234001}

\bibitem[\protect\citeauthoryear{{Vernstrom}, {Gaensler}, {Brown}, {Lenc}  \&
  {Norris}}{{Vernstrom} et~al.}{2017}]{2017MNRAS.467.4914V}
{Vernstrom} T.,  {Gaensler} B.~M.,  {Brown} S.,  {Lenc} E.,   {Norris} R.~P.,
  2017, \mn@doi [\mnras] {10.1093/mnras/stx424}, \href
  {https://ui.adsabs.harvard.edu/abs/2017MNRAS.467.4914V} {467, 4914}

\bibitem[\protect\citeauthoryear{{Vernstrom}, {Gaensler}, {Vacca}, {Farnes},
  {Haverkorn}  \& {O'Sullivan}}{{Vernstrom} et~al.}{2018}]{2018MNRAS.475.1736V}
{Vernstrom} T.,  {Gaensler} B.~M.,  {Vacca} V.,  {Farnes} J.~S.,  {Haverkorn}
  M.,   {O'Sullivan} S.~P.,  2018, \mn@doi [\mnras] {10.1093/mnras/stx3191},
  \href {https://ui.adsabs.harvard.edu/abs/2018MNRAS.475.1736V} {475, 1736}

\bibitem[\protect\citeauthoryear{{Vernstrom}, {Gaensler}, {Rudnick}  \&
  {Andernach}}{{Vernstrom} et~al.}{2019}]{2019ApJ...878...92V}
{Vernstrom} T.,  {Gaensler} B.~M.,  {Rudnick} L.,   {Andernach} H.,  2019,
  \mn@doi [\apj] {10.3847/1538-4357/ab1f83}, \href
  {https://ui.adsabs.harvard.edu/abs/2019ApJ...878...92V} {878, 92}

\bibitem[\protect\citeauthoryear{{Vernstrom}, {Heald}, {Vazza}, {Galvin},
  {West}, {Locatelli}, {Fornengo}  \& {Pinetti}}{{Vernstrom}
  et~al.}{2021}]{2021MNRAS.505.4178V}
{Vernstrom} T.,  {Heald} G.,  {Vazza} F.,  {Galvin} T.~J.,  {West} J.~L.,
  {Locatelli} N.,  {Fornengo} N.,   {Pinetti} E.,  2021, \mn@doi [\mnras]
  {10.1093/mnras/stab1301}, \href
  {https://ui.adsabs.harvard.edu/abs/2021MNRAS.505.4178V} {505, 4178}

\bibitem[\protect\citeauthoryear{{Welter}, {Perry}  \& {Kronberg}}{{Welter}
  et~al.}{1984}]{1984ApJ...279...19W}
{Welter} G.~L.,  {Perry} J.~J.,   {Kronberg} P.~P.,  1984, \mn@doi [\apj]
  {10.1086/161862}, \href
  {https://ui.adsabs.harvard.edu/abs/1984ApJ...279...19W} {279, 19}

\bibitem[\protect\citeauthoryear{{Wen} \& {Han}}{{Wen} \&
  {Han}}{2015}]{2015ApJ...807..178W}
{Wen} Z.~L.,  {Han} J.~L.,  2015, \mn@doi [\apj] {10.1088/0004-637X/807/2/178},
  \href {https://ui.adsabs.harvard.edu/abs/2015ApJ...807..178W} {807, 178}

\bibitem[\protect\citeauthoryear{{Wen}, {Han}  \& {Liu}}{{Wen}
  et~al.}{2012}]{2012ApJS..199...34W}
{Wen} Z.~L.,  {Han} J.~L.,   {Liu} F.~S.,  2012, \mn@doi [\apjs]
  {10.1088/0067-0049/199/2/34}, \href
  {https://ui.adsabs.harvard.edu/abs/2012ApJS..199...34W} {199, 34}

\bibitem[\protect\citeauthoryear{{Xu} \& {Han}}{{Xu} \&
  {Han}}{2021}]{2021arXiv211201763X}
{Xu} J.,  {Han} J.~L.,  2021, arXiv e-prints, \href
  {https://ui.adsabs.harvard.edu/abs/2021arXiv211201763X} {p. arXiv:2112.01763}

\bibitem[\protect\citeauthoryear{{You}, {Han}  \& {Chen}}{{You}
  et~al.}{2003}]{2003AcASn..44S.155Y}
{You} X.~P.,  {Han} J.~L.,   {Chen} Y.,  2003, Acta Astronomica Sinica, \href
  {https://ui.adsabs.harvard.edu/abs/2003AcASn..44S.155Y} {44, 155}

\bibitem[\protect\citeauthoryear{{Zonca}, {Singer}, {Lenz}, {Reinecke},
  {Rosset}, {Hivon}  \& {Gorski}}{{Zonca} et~al.}{2019}]{2019JOSS....4.1298Z}
{Zonca} A.,  {Singer} L.,  {Lenz} D.,  {Reinecke} M.,  {Rosset} C.,  {Hivon}
  E.,   {Gorski} K.,  2019, \mn@doi [The Journal of Open Source Software]
  {10.21105/joss.01298}, \href
  {https://ui.adsabs.harvard.edu/abs/2019JOSS....4.1298Z} {4, 1298}

\bibitem[\protect\citeauthoryear{{van Haarlem} et~al.,}{{van Haarlem}
  et~al.}{2013}]{2013A&A...556A...2V}
{van Haarlem} M.~P.,  et~al., 2013, \mn@doi [\aap]
  {10.1051/0004-6361/201220873}, \href
  {https://ui.adsabs.harvard.edu/abs/2013A&A...556A...2V} {556, A2}

\makeatother
\end{thebibliography}




\appendix

\section{RRM evolution with redshift  of the LoTSS sample in redshift bins with equal-number of sources }
\label{app:RM_iso}

\begin{figure}
   \centering
    \includegraphics[width=\columnwidth]{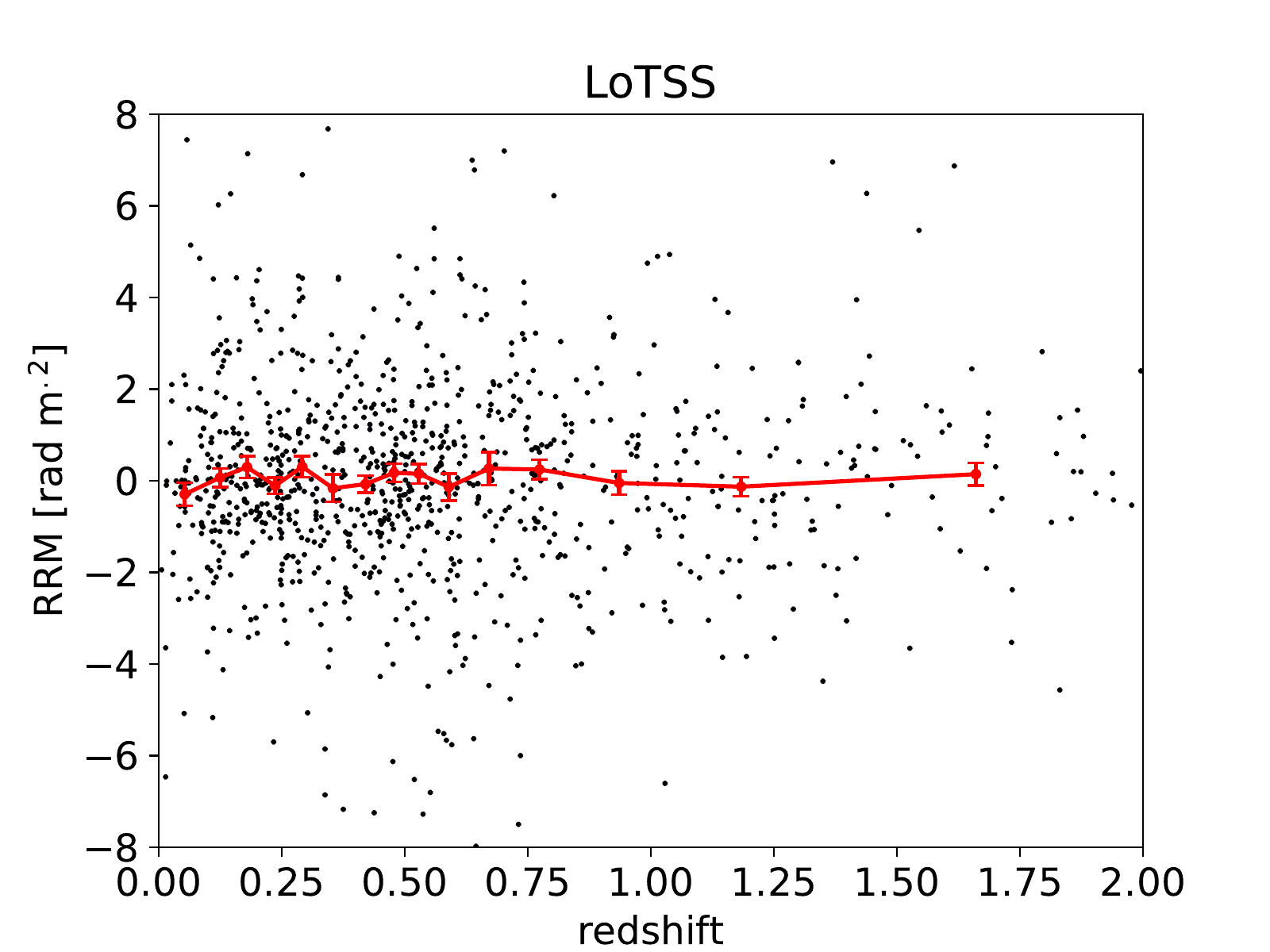}
    \includegraphics[width=\columnwidth]{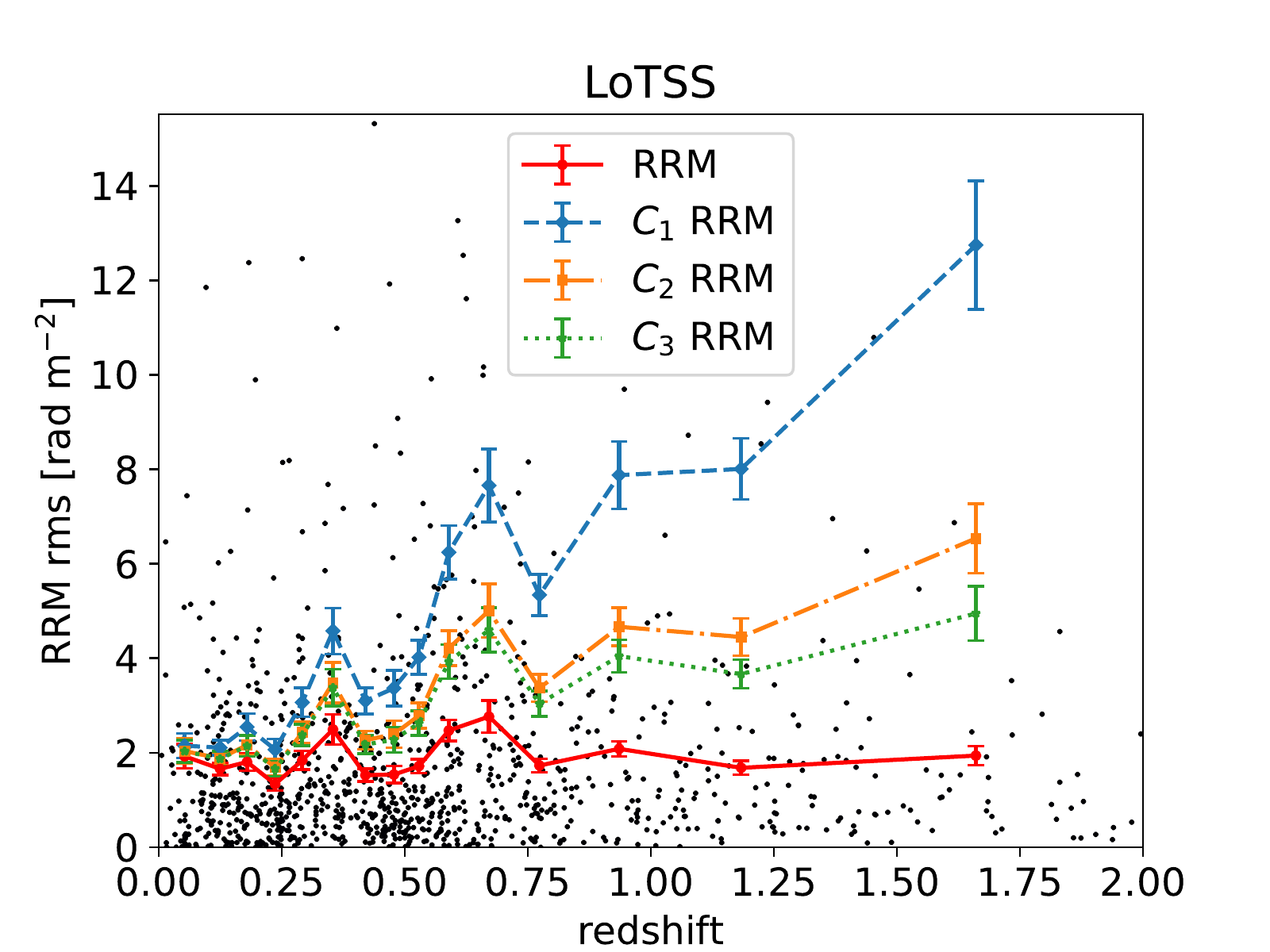}
   \caption{{\em Top}: As for Figure \ref{Fig:zmean20}, top panel, but the RRM mean is  in redshift bins with equal-number of sources (red solid line). {\em Bottom}: As for Figure \ref{Fig:zdisp20}, bottom panel, except the RRM rms deviation is in  redshift bins with equal-number of sources (red solid line). The number of bins is higher, redshift resolution is finer at low $z$ and coarser at high $z$.  The rest frame RRM$_0$ corrected for the redshift effect for the three models $C_x$ are also plotted.}
              \label{Fig:zmean_iso}%
              \label{Fig:zdisp_iso}%
    \end{figure}

We computed the RRM mean and rms deviation of the LoTSS sample in redshift bins with equal-number of sources, 64 each (Figure \ref{Fig:zmean_iso}). The redshift resolution is higher at low $z$ and coarser at high $z$, so this can give a better view of where the rest frame RRM$_0$ starts to increase. The uncorrected RRM is bumpy compared to the case with equal-width bins, but it is still flat with a linear-fit slope of $0.13 \pm 0.2$. All of  the  rest frame corrected RRM$_0$ are bumpy, but they show a clear increasing behaviour with $z$, starting from the low redshift end. A linear fit (Table \ref{tab:zfit_iso}) shows that the slope is again non-flat at a high confidence level: $\beta/\sigma_\beta = 11.2, 7.2, 5.1$ for the three correction models $C_1$ to $C_3$, with Student's t-test p--values of $p_t =  2.4\times10^{-8},  3.4\times 10^{-6}$, and  $  1.1\times10^{-4}$, respectively. That confirms a clear evolution of RRM$_0$ with $z$ (see also the Spearman tests in Table \ref{tab:zfit_iso}).

\begin{table*}
	\centering
	\caption{As for Table \ref{tab:zfit} except it is for the case of redshift bins with equal-numbers of sources.} 
	\label{tab:zfit_iso}
	\begin{tabular}{lcccccc}

        \hline 
        model & $\alpha $ &  $\beta$  &  $t = \beta/\sigma_\beta$  &  $p_t$  & $\rho$ & $p_\rho$ \\ 
           & [rad m$^{-2}$]  & [rad m$^{-2}$]  & & & & \\ 
       \hline 
        $C_1$ &  $1.24 \pm 0.42$  &  $6.64 \pm 0.59$  & 11.2  & $2.4\,\,10^{-8}$  & 0.94  &  $1.4\,\,10^{-7}$  \\ 
        $C_2$ &  $1.65 \pm 0.28$  &  $2.91 \pm 0.40$  & 7.2  & $3.4\,\,10^{-6}$  & 0.88  &  $1.4\,\,10^{-5}$ \\ 
        $C_3$ &  $1.87 \pm 0.27$  &  $1.97 \pm 0.39$  & 5.1  & $1.1\,\,10^{-4}$  & 0.87  &  $2.8\,\,10^{-5}$  \\ 
       \hline 

	\end{tabular}
\end{table*}

\section{Mean path length through a filament}
\label{app:len}

 The path length  along the line of sight through a filament of width $D$ and  inclination $\theta$ to the line of sight is
\begin{equation}
    l = \frac{D}{ \sin\theta}
\end{equation}
The mean path length averaged over  all filament orientations  thus is 
\begin{eqnarray}
    \bar{l} &=& \frac{1}{4\pi} \int_0^{2\pi}d\varphi \int_0^\pi \frac{D}{\sin\theta}\,\sin{\theta}\,d\theta \\
    &=& \frac{\pi}{2}\,D
\end{eqnarray}

\section{Filament magnetic field estimate with redshift bins with equal-number of sources}
\label{app:B_iso}
Following the procedure of Section \ref{sec:disc_cwf}, we repeated the estimate of the total  magnetic field  of individual filaments using the RRM rms measured in redshift bins with equal-number of sources  (Figure \ref{Fig:zdisp_iso}).  

\begin{figure}
   \centering
    \includegraphics[width=\columnwidth]{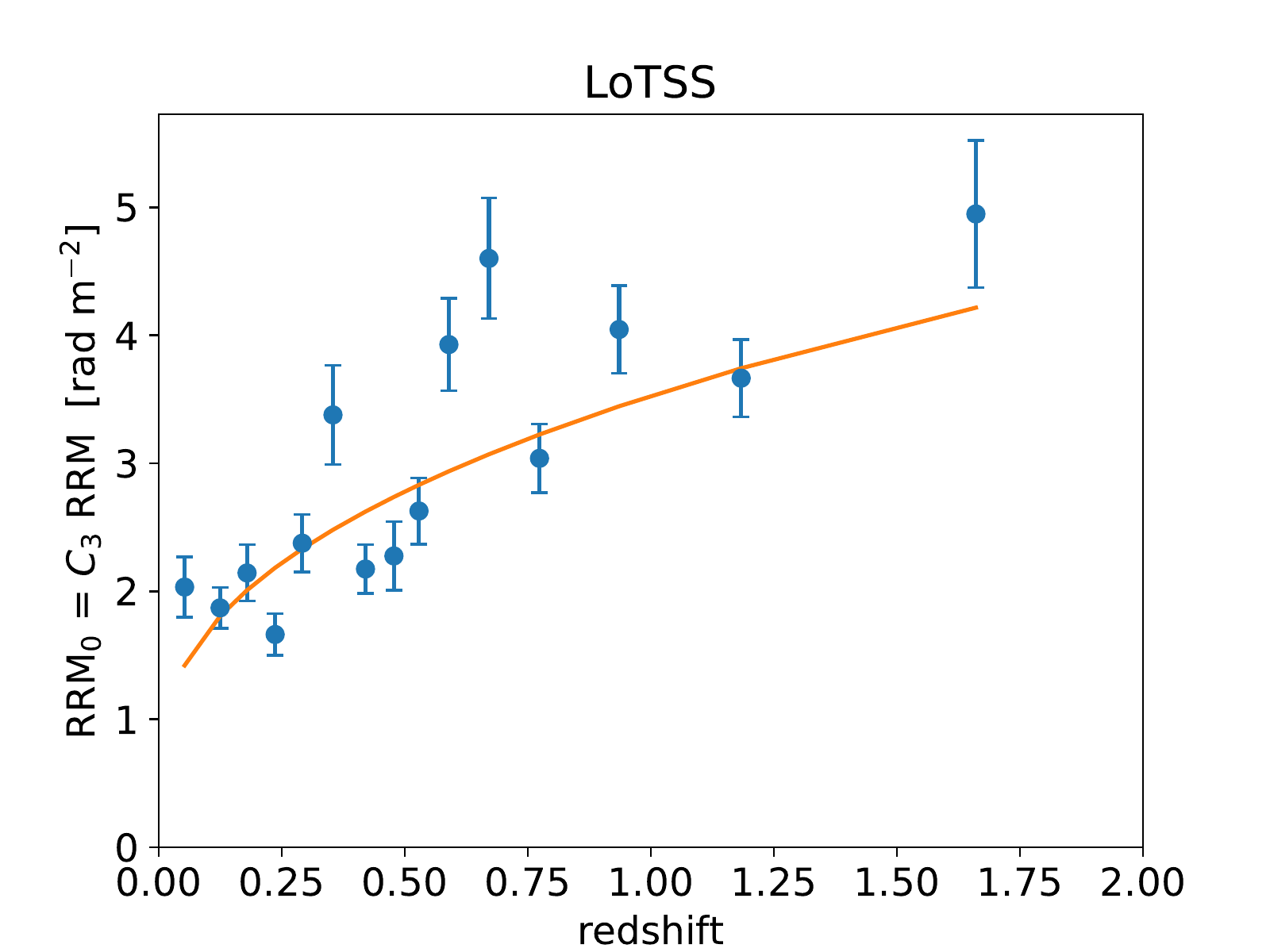}
   \caption{RRM$_0$ as a function of redshift of the LoTSS sample corrected for model $C_3$ (circles) in the case of redshift bins with equal-number of sources and best fit of the function of Equation (\ref{eq:rrmf}) (solid line).}
              \label{Fig:rrmfit_iso}%
    \end{figure}

The fit of Equation (\ref{eq:rrmf}) to the rest frequency RRM$_0$ rms is shown in Figure \ref{Fig:rrmfit_iso}. The resulting  filament RRM$_{0,f}$ is
\begin{equation}
    {\rm RRM}_{0,f} = 0.64 \pm 0.07  \,\, \rm rad\,m^{-2}
\end{equation}
Making  the same assumptions as for the main text we get a  filament total magnetic field of
\begin{equation}
    B_f = 29 \pm 3,\, {\rm nG},
\end{equation}
This is consistent with the result  of Section  \ref{sec:disc_cwf}, within the errors. 


\bsp	
\label{lastpage}
\end{document}